\RequirePackage{rotating}
\documentclass[useAMS,usenatbib]{mnras}

\usepackage{ulem}
\usepackage{color}
\usepackage{graphics,graphicx}
\usepackage{times} 
\usepackage{amssymb}
\usepackage{amsmath}
\usepackage{natbib}
\usepackage{lscape}
\usepackage{url}
\usepackage{multirow}
\usepackage{ctable}
\usepackage{threeparttable}
\usepackage[labelfont=bf,labelsep=period]{caption}
\newif\ifAMStwofonts
\AMStwofontstrue

\def\xmm{{\it XMM-Newton}}

\def\hst{{\it HST}}

\def\chandra{{\it Chandra}}
\def\swift{{\it Swift}}
\def\swiftng{{\it Neil Gehrels Swift Observatory}}

\def\epicpn{{EPIC-pn}}
\def\epicmos1{{EPIC-MOS1}}
\def\epicmos2{{EPIC-MOS2}}
\def\epicmos{{EPIC-MOS}}

\def\nustar{{\it NuSTAR}}

\def\gaia{{\it Gaia}}

\def\deg{$^{\circ}$}

\def\pcmsq{\hbox{$\rm\thinspace cm^{-2}$}}
\def\H0{{km~s$^{-1}$~Mpc$^{-1}$}}

\def\kev{\hbox{\rm keV}}

\def\ergpcmsqps{\hbox{$\rm\thinspace erg~cm^{-2}~s^{-1}$}}

\def\ergps{\hbox{erg~s$^{-1}$}}

\def\msun{\hbox{$M_{\odot}$}}

\def\addascaspec{\textsc{addascaspec}}

\def\flx2xsp{\textsc{flx2xsp}}
\def\sextractor{\textsc{sextractor}}

\def\nustardas{\textsc{nustardas}}
\def\nupipeline{\textsc{nupipeline}}
\def\nuproducts{\textsc{nuproducts}}

\def\sas{\textsc{sas}}
\def\xmmselect{\textsc{xmmselect}}

\def\epchain{\textsc{epchain}}
\def\emchain{\textsc{emchain}}

\def\rmfgen{\textsc{rmfgen}}
\def\arfgen{\textsc{arfgen}}

\def\ciao{\textsc{ciao}}
\def\chandrarepro{\textsc{chandra\_repro}}

\def\specextract{\textsc{specextract}}
\def\reprojectobs{\textsc{reproject\_obs}}
\def\wavdetect{\textsc{wavdetect}}
\def\wcsmatch{\textsc{wcsmatch}}
\def\wcsupdate{\textsc{wcsupdate}}

\def\hendrics{\textsc{hendrics}}

\def\chisq{{$\chi^{2}$}}

\def\xspec{\hbox{\small XSPEC}}

\def\simpl{\textsc{simpl}}

\def\tbabs{\textsc{tbabs}}

\def\diskbb{\textsc{diskbb}}
\def\diskpbb{\textsc{diskpbb}}

\def\kerrbb{\textsc{kerrbb}}

\def\cutoffpl{\textsc{cutoffpl}}

\def\eg{{\it e.g.}}

\def\ie{{\it i.e.~\/}}

\def\la{\mathrel{\hbox{\rlap{\hbox{\lower4pt\hbox{$\sim$}}}{\raise2pt\hbox{$<$}}}}}
\def\ga{\mathrel{\hbox{\rlap{\hbox{\lower4pt\hbox{$\sim$}}}{\raise2pt\hbox{$>$}}}}}

\def\d25{D$_{25}$}
\def\nh{{$N_{\rm H}$}}

\def\.25{0.25 keV\thinspace}

\def\kbol210{\rm $\kappa_{2-10}$}

\def\mbh{\rm $M_{\rm BH}$}

\def\rg{$R_{\rm{G}}$}

\def\rin{$R_{\rm in}$}

\def\rmag{$R_{\rm{M}}$}

\def\ngc{NGC\,7090}
\def\ulx3{NGC\,7090 ULX3}
\def\obsdate{2020-04-28}
\def\distance{9.5}
\def\namcsc{2CXO\,J213622.6$-$543234}
\def\namxmm{4XMM J213622.4$-$543233}

\title[A New ULX in NGC\,7090]{A New Transient Ultraluminous X-ray Source in NGC\,7090}

\author[D. J. Walton, et al.]
{\parbox{7.in}{D. J. Walton$^{1}$\thanks{E-mail: dwalton@ast.cam.ac.uk},
M. Heida$^{2}$, 
M. Bachetti$^{3}$, 
F. F\"urst$^{4}$, 
M. Brightman$^{5}$, 
H. Earnshaw$^{5}$, 
P. A. Evans$^{6}$, 
A. C. Fabian$^{1}$, 
B. W. Grefenstette$^{5}$, 
F. A. Harrison$^{5}$, 
G. L. Israel$^{7}$, 
G. B. Lansbury$^{2}$, 
M. J. Middleton$^{8}$, 
S. Pike$^{5}$, 
V. Rana$^{9}$, 
T. P. Roberts$^{10}$, 
G. A. Rodriguez Castillo$^{7}$, 
R. Salvaterra$^{11}$, 
X. Song$^{12}$, 
D. Stern$^{13}$ 
\\[0.25cm]
\footnotesize
$^{1}$ \it{Institute of Astronomy, University of Cambridge, Madingley Road, Cambridge CB3 0HA, UK} \\
$^{2}$ \it{European Southern Observatory, Karl-Schwarzschild-Stra$\beta$e 2, D-85748 Garching bei M\"unchen, Germany} \\
$^{3}$ \it{INAF-Osservatorio Astronomico di Cagliari, via della Scienza 5, I-09047 Selargius, Italy} \\
$^{4}$ \it{Quasar Science Resources for the European Space Agency (ESA), European Space Astronomy Centre (ESAC), Camino Bajo del Castillo s/n, 28692 Villanueva de la Cañada, Madrid, Spain} \\
$^{5}$ \it{Cahill Center for Astronomy and Astrophysics, California Institute of Technology, Pasadena, CA 91125, USA} \\
$^{6}$ \it{University of Leicester, School of Physics \& Astronomy, University Road, Leicester LE1 7RH, UK} \\
$^{7}$ \it{INAF-Osservatorio Astronomico di Roma, via Frascati 33, 00078 Monteporzio Catone, Italy} \\
$^{8}$ \it{Department of Physics and Astronomy, University of Southampton, Highfield, Southampton SO17 1BJ, UK} \\
$^{9}$ \it{Raman Research Institute, C. V. Raman Avenue, Sadashivanagar, Bangalore – 560080, India} \\
$^{10}$ \it{Centre for Extragalactic Astronomy, Durham University, Department of Physics, South Road, Durham DH1 3LE, UK} \\
$^{11}$ \it{INAF-Istituto di Astrofisica Spaziale e Fisica Cosmica di Milano, via A. Corti 12, 20133 Milano, Italy} \\
$^{12}$ \it{Jodrell Bank Centre for Astrophysics, Department of Physics and Astronomy, University of Manchester, Manchester M13 9PL, UK} \\
$^{13}$ \it{Jet Propulsion Laboratory, California Institute of Technology, Pasadena, CA 91109, USA}
}}
\date{}

\begin{document}
\pagerange{\pageref{firstpage}--\pageref{lastpage}}
\maketitle
\label{firstpage}

\begin{abstract}
We report on the discovery of a new, transient ultraluminous X-ray source (ULX) in the
galaxy NGC\,7090. This new ULX, which we refer to as \ulx3, was discovered via
monitoring with \swift\ during 2019--20, and to date has exhibited a peak luminosity of
$L_{\rm{X}} \sim 6 \times 10^{39}$\,\ergps. Archival searches show that, prior to its recent
transition into the ULX regime, ULX3 appeared to exhibit a fairly stable luminosity of
$L_{\rm{X}} \sim 10^{38}$\,\ergps. Such strong long-timescale variability may be
reminiscent of the small population of known ULX pulsars, although deep follow-up
observations with \xmm\ and \nustar\ do not reveal any robust X-ray pulsation signals.
Pulsations similar to those seen from known ULX pulsars cannot be completely excluded,
however, as the limit on the pulsed fraction of any signal that remains undetected in these
data is $\lesssim$20\%. The broadband spectrum from these observations is well
modelled with a simple thin disc model, consistent with sub-Eddington accretion, which
may instead imply a moderately large black hole accretor ($M_{\rm{BH}} \sim 40$\,\msun).
Similarly, though, more complex models consistent with the super-Eddington spectra seen
in other ULXs (and the known ULX pulsars) cannot be excluded given the limited
signal-to-noise of the available broadband data. The nature of the accretor powering this
new ULX therefore remains uncertain.
\end{abstract}


\begin{figure*}
\begin{center}
\rotatebox{0}{
{\includegraphics[width=425pt,trim={1.5cm 10cm 1.5cm 10cm},clip]{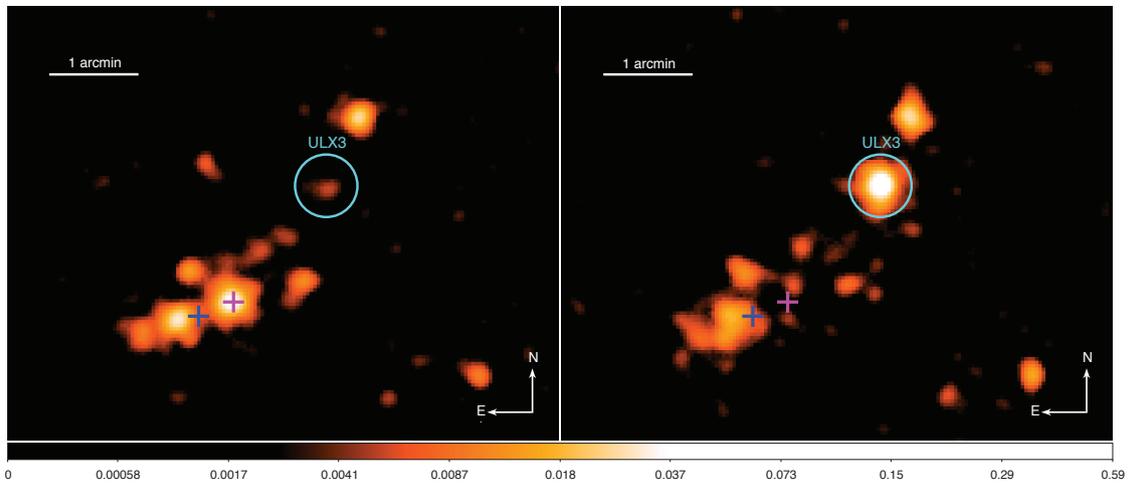}}
}
\end{center}
\vspace*{-0.3cm}
\caption{\swift\ XRT images (0.3--10.0\,keV) of \ngc, produced with the online XRT
pipeline (\citealt{Evans09}), taken over the lifetime of the mission prior to our monitoring
campaign (\textit{left}) and during our campaign (spread throughout 2019--2020;
\textit{right}). Both panels have been smoothed with a 4-pixel Gaussian for clarity and the
colour bar across the bottom -- common to both panels -- is in units of ct\,ks$^{-1}$. The
position of the newly discovered ULX3, which we report here, is shown with the cyan
circle. The blue cross (further to the east) shows the position of ULX1 (\eg\
\citealt{Song20}), seen in \xmm\ data but not obviously in either of the \swift\ images
(note that this is distinct from the persistent emission seen by \swift\ even further to the
east of ULX1), and the magenta cross (further to the west) shows the position of ULX2
(\eg\ \citealt{Liu19}), seen in the earlier \swift\ data but not during our more recent
campaign.
}
\label{fig_image}
\end{figure*}

\section{Introduction}

The population of Ultraluminous X-ray sources (ULXs) -- X-ray binaries which exhibit
luminosities in excess of $10^{39}$\,\ergps\  (see \citealt{Kaaret17rev} for a recent review)
-- is now generally understood to be primarily made up of compact objects accreting close
to or above their Eddington limits. This is driven by both spectroscopic and timing
observations. For sources that can be studied in detail, the broadband spectra provided
by \nustar\ (\citealt{NUSTAR}) are inconsistent with standard modes of sub-Eddington
accretion (\eg\ \citealt{Bachetti13, Rana15, Mukherjee15, Walton15hoII, Walton17hoIX}),
confirming the indications seen previously based on lower-energy data (\eg\
\citealt{Stobbart06, Gladstone09, Walton4517}), and are instead similar to the broad
expectation for super-Eddington accretion (emission from a hot and complex accretion
disc; \eg\ \citealt{Shakura73, Abram88}). In addition, powerful outflows have now been
observed in a number of ULXs through blueshifted atomic features (\citealt{Pinto16nat,
Pinto17, Pinto20, Walton16ufo, Kosec18}), as predicted for super-Eddington accretion.
Furthermore, a growing number of ULXs are now being identified as X-ray pulsars, which
must therefore be highly super-Eddington neutron star accretors (\citealt{Bachetti14nat,
Fuerst16p13, Israel17, Israel17p13, Carpano18, Sathyaprakash19, Rodriguez20};
see also \citealt{Brightman18}).\footnote{In addition to these more traditional ULX pulsars,
there are also a small number of neutron star Be/XRBs that occasionally and briefly peak
at $\sim$10$^{39}$\,\ergps\ during their largest outbursts, \eg\ A\,0538--66
(\citealt{Skinner82}), Swift\,J0243.6+6124
(\citealt{WilsonHodge18, Tao19}) and RX\,J0209.6--7427 (\citealt{Vasilopoulos20rx, Chandra20}). Although this is
distinct behaviour from the known ULX pulsars, which spend extended periods at highly
super-Eddington luminosities, these may be interesting objects in terms of connecting
ULX pulsars to the sub-Eddington X-ray pulsar population.} As the best local examples
of (relatively) sustained super-Eddington accretion, these sources may have significant
relevance for the growth of supermassive black holes in the early universe, given the
constraints placed by observations of $\sim$10$^{9}$\,\msun\ black holes when the
universe was less than a Gyr old (\eg\ \citealt{Mortlock11, Banados18nat}), so they are
potentially a key population to understand.

ULXs are known to be associated with recent star formation (\eg\ \citealt{Swartz11,
Mineo12, Lehmer19}), and are therefore generally thought of as high-mass X-ray binary
(HMXB) analogues. Indeed, where it has been possible to place robust constraints, either
via optical spectroscopy or X-ray timing, ULX stellar counterparts have generally been
found to be massive (\eg\ \citealt{Motch14nat, Bachetti14nat, Heida15, Heida16,
Heida19}). HMXBs are generally persistent sources, as the black hole or neutron star
feeds from the stellar wind of its massive companion (see \citealt{MartinezNunez17rev}
for a recent review). However, it is challenging to produce the high accretion rates needed
to explain the observed luminosities from ULXs purely via wind-fed accretion, and so
accretion via Roche-lobe overflow -- the primary mechanism by which low-mass X-ray
binaries (LMXBs) accrete (\eg\ \citealt{Verbunt93rev}) -- may be required despite their
apparent connection with the HMXB population (\eg\ \citealt{Fuerst18, ElMellah19}).
This may be related to specific periods in the evolution of the binary system, during which
mass can be transferred from the stellar companion on its thermal timescale (\eg\
\citealt{King01, Misra20}). Furthermore, as the archive of X-ray data continues to grow,
and along with it the number of nearby galaxies with multiple observing epochs, an
increasing number of new/transient ULXs are being reported (\eg\ \citealt{Soria12,
Middleton12, Middleton13nat, Pintore18, Pintore20, Earnshaw19, Earnshaw20,
vanHaaften19, Brightman20m51}). The high-amplitude long-term variability seen from
these systems is also generally difficult to explain in the context of persistent wind-fed
accretion, and is also more typically seen in LMXBs. However, even if the mass transfer
rate is relatively stable, such behaviour may still be possible for the magnetic neutron
stars in ULX pulsars if the propeller effect is invoked (\eg\ \citealt{Tsygankov16}). Indeed,
searching for sources that could be consistent with propeller transitions is a promising
means to identify ULX pulsar candidates among the broader ULX population (\eg\
\citealt{Earnshaw18, Song20}), so these highly variable ULXs are potentially of particular
interest.

Here we report on the discovery and characterisation of a new, transient ULX in \ngc\ ($z
= 0.00282$), utilizing observations with the \swiftng\ (hereafter \swift; \citealt{SWIFT}),
\xmm\ (\citealt{XMM}), \nustar\ (\citealt{NUSTAR}) and \chandra\ (\citealt{CHANDRA}).
Throughout this work, we assume a distance to \ngc\ of $D$ = \distance\,Mpc (based on
the tip of the red-giant branch method; \citealt{Karachentsev18}).

\section{\ulx3}

\ngc\ was previously known to host two highly variable/transient ULXs
(\citealt{WaltonULXcat, EarnshawULXcat, Liu19, Song20}). During a brief monitoring
programme targeting this galaxy with \swift\ throughout 2019--2020, primarily intended
to track the activity of the brighter of these two sources, we serendipitously discovered
that a third, previously unknown ULX had appeared in the XRT data at RA =
21$^{h}$36$^{m}$22.74$^{s}$, DEC = $-$54{\deg}32$'$33.8$''$ (see Section
\ref{sec_opt}), which we refer to as \ulx3\ (hereafter simply ULX3; see Figure
\ref{fig_image}). We therefore triggered a deep \xmm+\nustar\ target-of-opportunity
observation (epoch XN1) in order to investigate the spectral and temporal properties of
this new source. This observation was performed on \obsdate; further details are given
in Table \ref{tab_obs}. Both the \xmm\ and \nustar\ data were reduced following standard
procedures, as outlined below.

\begin{table}
  \caption{Details of the \xmm, \nustar\ and \chandra\ X-ray observations of \ngc\
  considered in this work.}
\vspace{-0.5cm}
\begin{center}
\begin{tabular}{c c c c c}
\hline
\hline
\\[-0.25cm]
Epoch & Mission & OBSID & Start & Good \\
& & & Date & Exposure\tmark[a] \\
\\[-0.3cm]
\hline
\hline
\\[-0.1cm]
\multicolumn{5}{c}{\textit{2020 Observations}} \\
\\[-0.2cm]
\multirow{2}{*}{XN1} & \nustar\ & 80501321002 & 2020-04-28 & 122 \\
\\[-0.3cm]
& \xmm\ & 0852050201 & 2020-04-29 & 82/100 \\
\\
\multicolumn{5}{c}{\textit{Archival Data}} \\
\\[-0.2cm]
X1 & \xmm\ & 0200230101 & 2004-04-08 & -- \\
\\[-0.3cm]
X2 & \xmm\ & 0200230201 & 2004-05-13 & 6/11 \\
\\[-0.3cm]
C1 & \chandra\ & 7060 & 2005-12-18 & 26 \\
\\[-0.3cm]
C2 & \chandra\ & 7252 & 2006-04-10 & 31 \\
\\[-0.3cm]
X3 & \xmm\ & 0503460101 & 2007-10-05 & 6/8 \\
\\[-0.2cm]
\hline
\hline
\\[-0.15cm]
\end{tabular}
\label{tab_obs}
\end{center}
\vspace{-0.3cm}
$^{a}$ Exposures are given in ks, and for \xmm\ are listed for the \epicpn/MOS detectors
after background flaring has been excised.
\end{table}

\subsection{Observations and Data Reduction}
\label{sec_red}

The \nustar\ data were reduced with the \nustar\ Data Analysis Software (\nustardas)
v1.9.2, and \nustar\ calibration database v20190627. First the unfiltered event files for
both focal plane modules (FPMA and FPMB) were cleaned with \nupipeline, using the
standard depth correction to reduce the internal high-energy background and excluding
passages through the South Atlantic Anomaly. The data for both modules were
corrected to the solar barycentre using the DE200 solar ephemeris. Source products
and their associated instrumental response files were then extracted for each module
using circular regions of radius 30$''$ with \nuproducts. For all the datasets considered
here, their associated backgrounds were estimated from larger regions of blank sky on
the same chip as ULX3. In order to maximise the signal-to-noise (S/N), we use both
the standard `science' data (mode 1) and the `spacecraft science' data (mode 6; see
\citealt{Walton16cyg}); the mode 6 data provide $\sim$14\% of the total \nustar\
exposure quoted in Table \ref{tab_obs}. Finally, given the moderate signal-to-noise of
the data for the individual focal plane modules, we combined their spectra using
\addascaspec; we also note that this observation was not affected by the recently
identified tear in the thermal blanket (\citealt{NuSTARmli}). The \nustar\ data provide
constraints up to $\sim$20--25\,keV in this case; above these energies there is no
significant detection of ULX3 above the background level.

The \xmm\ data were reduced using the \xmm\ Science Analysis System (\sas\ v18.0.0).
Raw observation files were cleaned using \epchain\ and \emchain\ for the \epicpn\ and
\epicmos\ detectors, respectively (\citealt{XMM_PN, XMM_MOS}). The cleaned event
files for \epicpn, which has the best time resolution of the EPIC detectors (73.4\,ms in
the full frame mode used here), were corrected to the solar barycentre using the
DE200 solar ephemeris, similar to the \nustar\ data. Source products were extracted
from the cleaned event files with \xmmselect\ from a circular region of radius 30$''$;
periods of high background were removed as standard, but only occur at the very end
of the observation. As recommended, we only considered single and double patterned
events for \epicpn\ ({\small PATTERN}\,$\leq$\,4) and single to quadruple patterned
events for \epicmos\ ({\small PATTERN}\,$\leq$\,12). The instrumental response files
were generated using \arfgen\ and \rmfgen\ for each of the EPIC detectors. After
performing the reduction separately for the two \epicmos\ units, we also combined
these data using \addascaspec.

\begin{figure}
\begin{center}
\hspace*{-0.35cm}
\rotatebox{0}{
{\includegraphics[width=235pt]{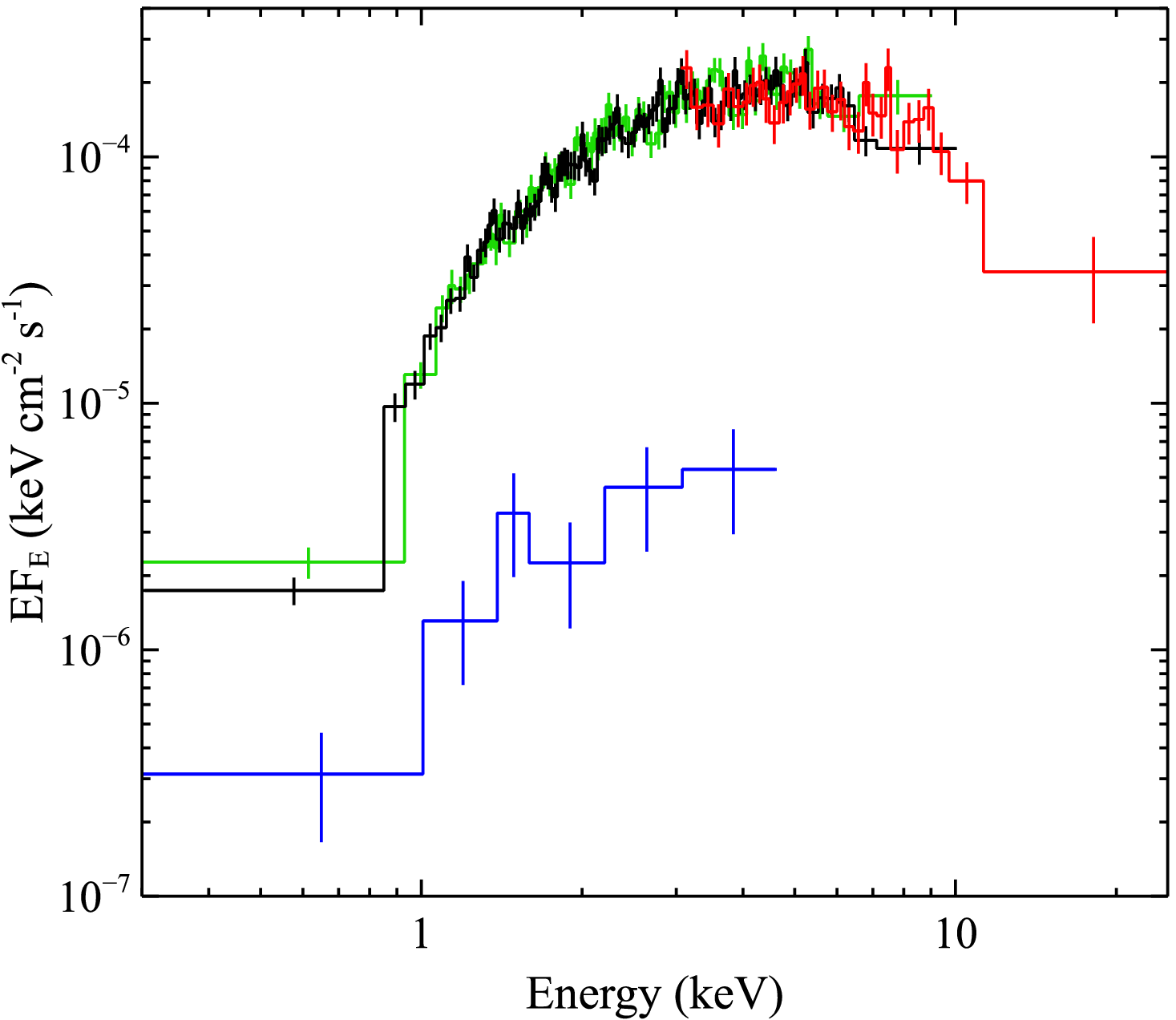}}
}
\end{center}
\vspace*{-0.3cm}
\caption{The broadband \xmm+\nustar\ X-ray spectrum of \ulx3\ taken in 2020 (epoch
XN1). The data have been unfolded through a model that is constant with energy, and
rebinned for visual clarity. For epoch XN1, the \xmm\ EPIC-pn and EPIC-MOS data are
shown in black and green, respectively, and the \nustar\ FPM data are shown in red. We
also show a comparison with one of the low-flux archival observations, epoch C1, with
the ACIS-S data in blue.
}
\label{fig_spec}
\end{figure}

\subsection{Spectroscopy}
\label{sec_spec}

The broadband X-ray spectrum extracted from the 2020 \xmm+\nustar\ dataset is shown
in Figure \ref{fig_spec}. The data have a distinctly thermal appearance, peaking at 
$\sim$5\,keV before falling away with a steep spectrum at higher energies, qualitatively
similar to the other ULXs observed by \nustar\ to date (\eg\ \citealt{Walton18p13,
Walton18ulxBB}). The flux from ULX3 is stable throughout the observation (see Section
\ref{sec_pulse}), so we apply some simple continuum models to the time-averaged
broadband data using \xspec\ (v12.10.1s; \citealt{xspec}). All the individual datasets from
epoch XN1 are rebinned to a minimum S/N of 5 per energy bin, and the data are fit by
reducing the \chisq\ statistic. As is standard, we allow for normalisation constants to vary
between the datasets to account for residual cross-calibration issues, fixing \epicpn\ to
unity; for the models that successfully fit the data, these are all within normal ranges
(\citealt{NUSTARcal}). Neutral absorption is modeled with \tbabs, combining the
cross-sections of \cite{Verner96} and the solar abundance set of \cite{tbabs}. In all of the
models considered here, we include absorption component that corresponds to the
Galactic column ($2 \times 10^{20}$\,\pcmsq; \citealt{NH2016}), and also separately
allow for absorption local to the source that is free to vary.

\begin{figure}
\begin{center}
\hspace*{-0.35cm}
\rotatebox{0}{
{\includegraphics[width=235pt]{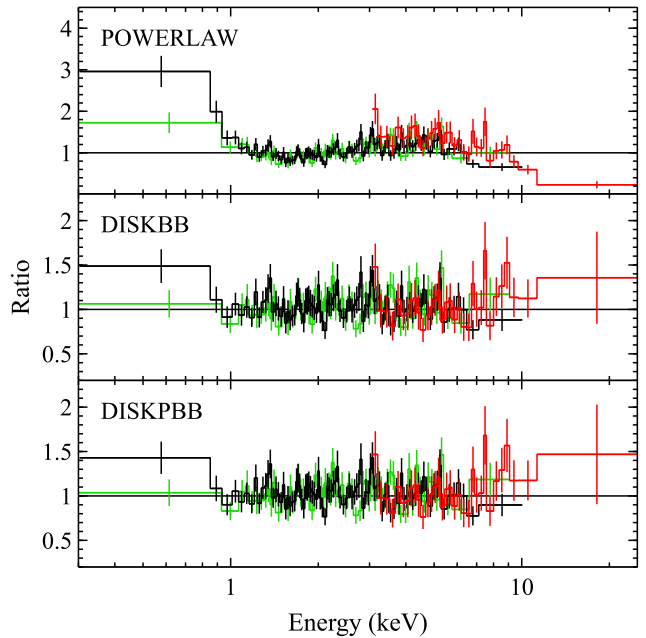}}
}
\end{center}
\vspace*{-0.3cm}
\caption{Data/model ratios for the simple continuum models applied to the broadband
dataset (epoch XN1) for ULX3 (Section \ref{sec_spec}). The colours have the same
meanings as in Figure \ref{fig_spec}.
}
\label{fig_ratio}
\end{figure}

We first apply an absorbed powerlaw continuum. However, this provides a very poor fit
to the data, with \chisq\ = 635 for 326 degrees of freedom (DoF). Given the thermal
appearance of the data, we therefore also try a simple accretion disc model, using
\diskbb\ (\citealt{diskbb}). This assumes a standard thin accretion disc, based on the
model of \cite{Shakura73}. In contrast to the powerlaw continuum, this simple model
actually provides a very good fit to the data, with \chisq/DoF = 341/326. Given the
evidence that most ULXs are likely super-Eddington accretors, we also fit the \diskpbb\
accretion disc model (\citealt{diskpbb}). This allows the radial temperature index, $p$, to
vary as an additional free parameter (defined such that $T(r) \propto r^{-p}$), and is often
used to approximate super-Eddington accretion discs. These are expected to have a
significant scale height and, as discussed by \cite{Abram88}, should be characterised by
$p \sim 0.5$ (as opposed to the thin disc solution, for which $p = 0.75$). Although
the \diskpbb\ model also fits the data very well, with \chisq/DoF = 341/325, this offers a
negligible improvement over the simpler \diskbb\ model despite the extra free parameter
($\Delta\chi^{2} \sim 0.5$). Indeed, when allowed to vary, we find that $p$ is consistent
with the standard thin disc solution in this case. The results obtained with these models
are given in Table \ref{tab_XN1}, and the data/model ratios for each of these fits are
shown in Figure \ref{fig_ratio}.

\begin{table}
  \caption{Best fit parameters obtained for the simple continuum models applied
  to the broadband spectrum observed from ULX3 (epoch XN1)}
\begin{center}
\begin{tabular}{c c c c c c}
\hline
\hline
\\[-0.25cm]
Model: & Powerlaw & \diskbb\ & \diskpbb \\
\\[-0.3cm]
\hline
\hline
\\[-0.2cm]
\nh\ ($10^{21}$\,cm$^{-2}$) & $16.5 \pm 0.1$ & $7.9\pm0.5$ & $7.4\pm1.2$ \\
\\[-0.275cm]
$\Gamma$ & $2.05\pm0.06$ & -- & -- \\
\\[-0.275cm]
$T_{\rm{in}}$ (\kev) & -- & $1.87 \pm 0.07$ & $1.8^{+0.2}_{-0.1}$ \\
\\[-0.275cm]
$p$ & -- & -- & $0.8 \pm 0.1$ \\
\\[-0.275cm]
Norm ($10^{-4})$ & $1.8^{+0.1}_{-0.2}$ & $23 \pm 3$ & $28^{+21}_{-11}$ \\
\\[-0.2cm]
\hline
\\[-0.25cm]
$\chi^{2}$/DoF & 635/326 & 341/326 & 341/325 \\
\\[-0.25cm]
\hline
\hline
\end{tabular}
\label{tab_XN1}
\end{center}
\end{table}

Following previous work on ULX spectroscopy, we try a variety of more complex models, 
but find that these are not required by the data from epoch XN1. Below 10\,keV, high
S/N ULX spectra typically require the presence of two thermal disc-like components,
typically with temperatures $\sim$0.3 and $\sim$3\,keV (\eg\ \citealt{Stobbart06,
Gladstone09}). The hotter component (also seen here) is likely associated with the
innermost accretion flow, while the lower temperature component is potentially
associated with either the outer accretion flow or the photosphere of a strong,
super-Eddington wind (\eg\ \citealt{King03}). However, adding a second,
low-temperature thermal component to the baseline \diskbb\ model only improves the fit
by $\Delta\chi^{2} = 3$ for two more free parameters. It is worth noting, though, that the
line-of-sight absorption column towards ULX3 is quite high in this case, and so the
presence of an additional low-temperature emission component could easily be masked
by absorption.

At higher energies, all of the ULXs observed by \nustar\ with good S/N to date show
evidence for additional continuum emission above $\sim$10\,keV when the lower-energy
data are fit with accretion disc models (\eg\ \citealt{Walton14hoIX, Walton15hoII,
Walton18ulxBB, Mukherjee15, Fuerst17ngc5907}). In the case of the ULX pulsars, this
emission is known to be associated with the central, magnetically-channelled accretion
columns (\citealt{Brightman16m82a, Walton18crsf, Walton18p13}), while for
non-magnetic neutron star/black hole ULXs this emission would presumably arise from a
Compton-scattering corona. Adding an additional high-energy component to represent
either of these possibilities -- using the \simpl\ convolution model (\citealt{simpl})
for the powerlaw emission from a corona and the template for ULX pulsar accretion
columns adopted in \citet[][a \cutoffpl\ component with $\Gamma = 0.59$ and
$E_{\rm{cut}} = 7.9$\,keV]{Walton20} -- also results in a negligible improvement in
the fit, with $\Delta\chi^{2} = 1$ in both cases. However, the high-energy ($E > 10$\,keV)
S/N is significantly lower in this case than for any of the sources in the \nustar\ sample
discussed in \cite{Walton18ulxBB}, where evidence for the extra high-energy component
was ubiquitously seen. In the case of the \simpl\ model, we find an upper limit on the
scattered fraction (which acts as an effective normalisation for the powerlaw flux) of
$f_{\rm{sc}} < 32\%$, which is not dissimilar to the values seen in other ULXs when fit
with a similar model (typical values are a few 10s of \%, \eg\ \citealt{Walton15hoII}).

For the \diskbb\ model, which provides the simplest explanation of the broadband data,
we compute the observed flux in the 0.3--10\,keV band, for comparison with archival
datasets (Section \ref{sec_archive}), and absorption corrected luminosities in the
0.3--10.0\,keV and 0.3--30.0\,keV bands. We find $F_{\rm{obs,0.3-10}} = (4.4 \pm 0.1)
\times 10^{-13}$\,\ergpcmsqps, $L_{\rm{int,0.3-10}} = (6.0 \pm 0.2) \times
10^{39}$\,\ergps\ and $L_{\rm{int,0.3-30}} = (6.4 \pm 0.2) \times 10^{39}$\,\ergps\
(assuming isotropic emission). Similar to other ULXs, the vast majority of the broadband
flux is emitted below 10\,keV.

\subsection{Timing Analysis}
\label{sec_pulse}

\begin{figure}
\begin{center}
\hspace*{-0.35cm}
\rotatebox{0}{
{\includegraphics[width=235pt]{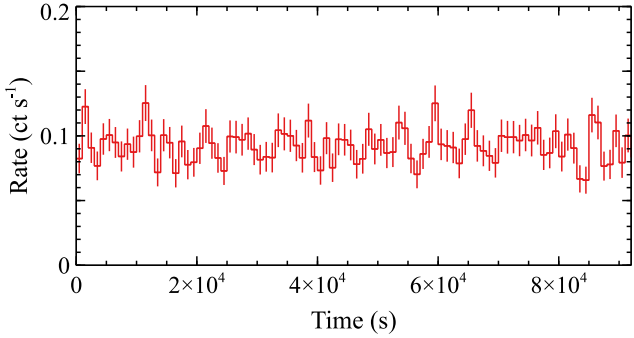}}
}
\end{center}
\vspace*{-0.3cm}
\caption{The 0.3--10.0\,keV \epicpn\ lightcurve of \ulx3\ during epoch XN1, shown with
1\,ks time bins.
}
\label{fig_shortlc}
\end{figure}

The X-ray lightcurve from the \epicpn\ detector for epoch XN1 is shown in Figure
\ref{fig_shortlc}; overall, the source flux appears to be relatively stable during
these observations. We therefore limit our variability analysis to a search for
pulsations in order to explore the possibility that ULX3 is a new member of the ULX
pulsar population.

We use several strategies to look for pulsations -- ranging from general pulsation
searches to deeper searches based on the properties of known ULX pulsars --
utilizing the pulsar timing tools included in \hendrics\
(\citealt{bachettiHENDRICSHighENergy2018}). We focus primarily on the data from
the \epicpn\ camera, as this has both the highest count rates and the best temporal
resolution of the EPIC detectors, and investigate candidate pulsations below a
Nyquist frequency corresponding to the 73-ms frame time of \epicpn. Our goal is to
find pulsations whose frequency might be changing quickly during our observation
due to the intrinsic spin up/down (\citealt{Israel17}) or to orbital motion (\eg\
\citealt{Bachetti14nat, Rodriguez20}). The first effect should give rise to an
approximately constant acceleration (\ie\ a linear change) of the pulse frequency
during the observation. 

For this analysis, we maximise the S/N by running a Fourier-space accelerated search
(\citealt{PRESTOaccel}) on the whole lightcurve using the \textsc{HENaccelsearch}
tool. The range in frequency derivatives we search formally depends on the frequency
being searched, but at a central frequency of 1\,Hz we search a range of $\pm$ 5
$\times 10^{-8}$\,Hz/s with a resolution of $2.5 \times 10^{-10}$\,Hz/s. This task uses
standard ``detection levels'' based on the $\chi^{2}$ distribution (\citealt{Leahy83b}).
To account for the different sensitivity response across a given frequency bin (which
leads to a drop of sensitivity as the pulsation frequency departs from the central
frequency and approaches the edge of the bin), the tool initially considers all
frequencies with variability powers in excess of 36\% of the power level corresponding
to a 1\% false-alarm probability as ``candidate'' pulsations. These candidates are then
investigated with more sensitive $Z^2_n$ search \citep{Buccheri83}. Additionally, we
also run the accelerated search on shorter, overlapping intervals of the light curve
(down to 50\,ks) to look for non-linear frequency changes (from orbital motion) and
transient pulsations. Motivated by the generally hard pulsed spectra in ULX pulsars
(\eg\ \citealt{Brightman16m82a, Walton18p13}), we repeat these searches both on
the full event lists and in the 2.5--12\,keV energy range. Since most ULX pulsars have
quasi-sinusoidal profiles, we do not perform any harmonic summing in the power
density spectra calculated by \textsc{HENaccelsearch} and we limit the number $n$
of harmonics in the $Z^2_n$ searches to 2. 

There are a few candidate frequencies (at 0.88 and 4.36\,Hz for the 2.5--12\,keV
bandpass and 6.4\,Hz for the full dataset) that slightly exceeded the nominal detection
level. Unfortunately, these timescales are all too fast for independent verification with
the \epicmos\ detectors, and the \nustar\ data are also unable to independently
confirm whether any of them represent a real signal (although the \nustar\ limits are
less stringent owing to its lower count rates). Therefore, given the lack of consistency
between the full band and the harder band, and the lack of independent verification
with any other available detector, we do not consider any of these potential signals to
be robust, \textit{bona-fide} detections. Nevertheless, we report them in case one of
these frequencies is seen again in future observations with similar significance (with a
blind search), as this would potentially transform one of these marginal cases into a
more believable pulsation signal.

Given the lack of a robust pulsation detection, we calculate the upper limit on the
pulsed fraction any pulsed signal could have using \textsc{HENzn2vspf}, focusing on
the 0.01--7\,Hz frequency range. \textsc{HENzn2vspf} simulates datasets using the
same GTIs and total number of events in each GTI as seen in the real data, and uses
rejection sampling to modulate the events with stronger and stronger pulsations. For
each simulated dataset, the tool calculates the $Z^2_2$ statistics and produces a
$Z^2_2$ versus pulsed fraction plot. We simulate 100 datasets with increasing
pulsed fraction, and determine the point where the $Z^2_2$ reaches $\sim$40. This
roughly corresponds to a 3$\sigma$ statistical detection, and thus gives the
equivalent upper limit on the pulsed fraction permitted by the real data. We find limits
on the pulsed fraction of $\sim20$\% for the full dataset and $\sim40$\% for the
energy-selected or time-selected intervals. We note that these values are similar to
the pulsed fractions inferred for the marginal pulsation candidates listed above,
confirming again that they should be treated with caution.

\begin{figure*}
\begin{center}
\hspace*{-0.15cm}
\rotatebox{0}{
{\includegraphics[width=510pt]{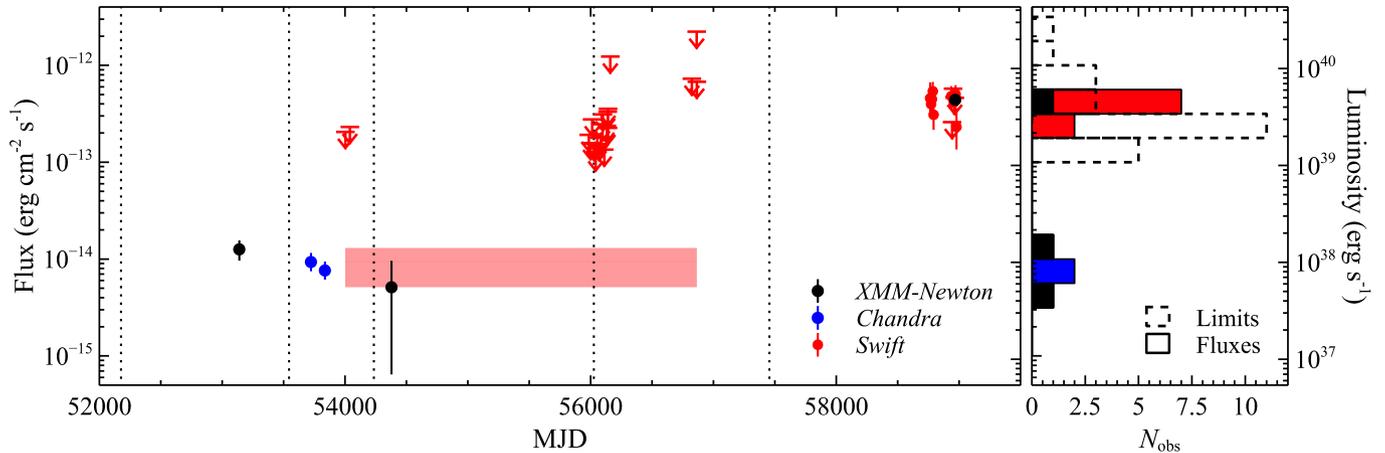}}
}
\end{center}
\vspace*{-0.3cm}
\caption{The long-term 0.3--10\,keV X-ray lightcurve of ULX3 (\textit{left panel}),
and the corresponding flux distribution (\textit{right panel}). Data from \xmm,
\chandra\ and \swift\ are shown in black, blue and red, respectively. The pink shaded
region corresponds to the average flux seen stacking all of the archival \swift\
observations. ULX3 shows both high- and low-flux states, separated by more than an
order of magnitude, similar to known ULX pulsars. The timing of the \textit{HST}
observations considered in Section \ref{sec_opt} are indicated with vertical dotted
lines.
}
\label{fig_longlc}
\end{figure*}

\section{Archival Data \& Long-Term Variability}
\label{sec_archive}

In addition to the new \xmm+\nustar\ observation presented here, there are also three
archival observations of \ngc\ with \xmm, and two with \chandra, which we also consider
to provide contextual information over a longer timescale; these are spread over the
period 2004--2007 (see Table \ref{tab_obs}). The two \chandra\ observations, both taken
with the ACIS-S detector (\citealt{CHANDRA_ACIS}), were reduced with \ciao\ v4.11
and its associated calibration files. Cleaned event files were generated with the
\chandrarepro\ script. ULX3 is clearly detected in both \chandra\
observations,\footnote{\label{fn_cats}The low-flux detections of ULX3 are included in
the CSC2 and 4XMM catalogues (\citealt{CSC2temp, 4XMM}) under the identifiers
\namcsc\ and \namxmm.} albeit at much lower fluxes, and so source spectra and
instrumental response files were extracted with the \specextract\ script from circular
regions of radius $2''$. For the first \xmm\ observation (epoch X1), the entire exposure
suffered from severe background flaring, so we do not make any use of these data.
ULX3 is clearly detected in the second observation (epoch X2),$^{\ref{fn_cats}}$ and
appears to be marginally detected in the third (epoch X3). As with the \chandra\ data,
the source fluxes are significantly lower than epoch XN1, so we extract source spectra
largely as outlined above, but using a smaller source region (radius 15$''$). The only
difference is that for epoch X3, the position of ULX3 fell on a bad row for the \epicmos1\
detector, so in this case we only utilize the data from the \epicpn\ and \epicmos2\
detectors.

Of these archival datasets, epoch C1 has the highest S/N, and so we show this in
comparison to the high-flux data in Figure \ref{fig_spec}. Over the more limited
bandpass covered, the observed spectrum is still quite hard, although this is not
surprising given the reasonably substantial absorption column inferred from the epoch
XN1 data. To model these low-flux data, given the low S/N, we group them to 1 count
per bin and fit them by reducing the Cash statistic (\citealt{cstat}). We also fix the
level of absorption to that found previously (\ie \nh\ = $7.9 \times 10^{21}$\,\pcmsq).
For both powerlaw and accretion disc continuum models (assuming a thin disc for
simplicity), we find that the data from epochs C1, C2, X2 and X3 can all be fit with a
common spectral shape; with the former we find $\Gamma = 2.2 \pm 0.4$ and with the
latter we find $T_{\rm{in}} = 0.9 \pm 0.2$\,keV. We use the latter to compute observed
fluxes in the 0.3--10\,keV band for each of these epochs (the fluxes computed with the
powerlaw continuum are systematically $\sim$20\% higher, but are ultimately in good
statistical agreement with the \diskbb\ fluxes). These are typically
$\sim$10$^{-14}$\,\ergpcmsqps\ for each of these epochs, corresponding to
luminosities of $\sim$10$^{38}$\,\ergps\ (again, assuming isotropic emission).
Interestingly, with the \diskbb\ model, we find the normalisations are all consistent at
the 90\% level, and the best-fit values are all in the range 1--2 $\times$ 10$^{-3}$,
broadly similar to the value seen from epoch XN1. Indeed, if we fit the low-flux data
with a common normalisation, instead of a common disc temperature, we find that the
normalisation is $1.4^{+2.1}_{-0.9} \times 10^{-3}$, while the best-fit disc temperatures
vary between $T_{\rm{in}} \sim 0.7-0.9$\,keV. The \diskbb\ normalisation is proportional
to $(R_{\rm{in}}/D)^2 \cos(i)$, where \rin\ is the inner radius of the disc, $D$ is the
distance to the source and $i$ is the disc inclination (see below), and so the data are
formally consistent with being dominated by a standard thin disc which has a constant
inner radius across all of the observing epochs to date (although we stress that the
uncertainties are large for the low-flux data).


We also extract the \swift\ 0.3--10.0\,keV lightcurve from the online XRT pipeline
(\citealt{Evans09}) in order to build a more complete picture of the flux history of ULX3.
In addition to the 12 observations taken in 2019/20 as part of our recent program
(typically $\sim$3--4\,ks exposures), \swift\ has also observed NGC\,7090 on a further
21 occasions (typically $\sim$2--4\,ks exposures) between 2006--2014 (spanning MJD
$\sim$ 54000--57000). Although there appears to be a weak detection in the integrated
XRT image from all of the observations that were taken prior to our more recent
campaign (see Figure \ref{fig_image}), ULX3 is not significantly detected in any of
these individual \swift\ observations. The XRT count rates (and their limits) are
converted to fluxes for comparison with the other datasets assuming the spectral forms
discussed above; for our recent monitoring campaign we use the spectral form found
for epoch XN1, while for the archival data we use the spectral form found for epochs
C1, C2, X2 and X3. Then, combining the 0.3--10.0\,keV flux from epoch XN1 with all of
these fluxes, we construct the long-term lightcurve of ULX3 shown in Figure
\ref{fig_longlc}.

Although the coverage is quite sparse, the detections provided by \xmm, \chandra\ and
\swift\ show evidence for high-amplitude (more than an order of magnitude) long-term
variability that is consistent with a bi-modal flux distribution. To further test this potential
bi-modality, we also extract the spectrum for the weak detection seen in the integrated
archival \swift\ data in Figure \ref{fig_image} (\ie prior to MJD 58000). Modelling this
with the same spectral form as seen in the rest of the low-flux archival datasets (a
thin disc with $T_{\rm{in}} = 0.9$\,keV), again binning the data to 1 count per bin and
reducing the Cash statistic, we find an average flux of $\sim$10$^{-14}$\,\ergpcmsqps\
for these data, consistent with the rest of the low-flux epochs. Although we use all of
the \swift\ data taken prior to 2020, this is dominated by a dense period of monitoring
of \ngc\ around MJD $\sim$ 56000, $\sim$4-5 years after the last archival \xmm\
observation. This flux level does therefore appear to represent a fairly stable baseline
for ULX3 prior to the observations obtained in 2020.

\section{Optical Counterparts}
\label{sec_opt}

As \ngc\ has been observed on several occasions by the \textit{Hubble Space Telescope}
(\hst), we also perform a search for any optical counterparts to ULX3 (see Table
\ref{tab_hst} for a list of the observations used here). For this analysis, we
focus on the imaging data from \chandra, and extract a combined image from the two
\chandra\ observations using the \ciao\ task \reprojectobs.

\begin{table*}
  \caption{Details of the \hst\ observations covering the position of ULX3. Magnitudes of
  the candidate optical counterparts are apparent instrumental Vega magnitudes
  computed with {\sc Dolphot}.}
\begin{center}
\begin{tabular}{c c c c c c c}
\hline
\hline
\\[-0.25cm]
Inst & Prop ID & Obs date & Filter & Exp time (s) & Src 1 & Src 2 \\
\\[-0.3cm]
\hline
\hline
\\[-0.3cm]
 WFPC2	& 09042	& 2001-09-24	&  F450W	& 460  &  $24.73  \pm 0.18$	    & $>25.3	$ \\
 WFPC2	& 09042	& 2001-09-24	&  F814W	& 460  &  $22.60  \pm 0.06$	    & $>24.6	$ \\ 
 ACS	& 10416	& 2005-06-23	&  F625W	& 2508 &  $23.227 \pm 0.010$	& $26.8  \pm 0.1$	\\
 ACS	& 10416	& 2005-06-23	&  F658N	& 7496 &  $22.94  \pm 0.03$	    & $25.9  \pm 0.4$\\
 WFPC2	& 10889	& 2007-05-14/17 &  F814W	& 6000 &  $22.528 \pm 0.012$	& $24.41 \pm 0.05$	\\
 ACS	& 12546	& 2012-04-09	&  F606W	& 900  &  $23.466 \pm 0.014$	& $27.6  \pm 0.4$	\\
 ACS 	& 12546	& 2012-04-09	&  F814W	& 900  &  $22.480 \pm 0.013$	& $24.97 \pm 0.08$	\\
 WFC3	& 14095	& 2016-03-08	&  F110W	& 298  &  $21.833 \pm 0.018$	& $22.05 \pm 0.02$	\\
 WFC3	& 14095	& 2016-03-08	&  F128N	& 903  &  $21.88  \pm 0.03$	    & $22.11 \pm 0.03$	\\
 \\[-0.3cm]
\hline
\hline
\end{tabular}
\label{tab_hst}
\end{center}
\end{table*}

In order to register the various images to a common coordinate system, we produce
source lists for both \chandra\ and \hst\ using \wavdetect\ and \sextractor\ for \chandra\
and \hst, respectively (adopting fairly standard source detection thresholds in each case),
each of which are matched against the \gaia\ DR2 source
catalogue\footnote{Unfortunately there are not enough \chandra\ sources in the relatively
small field of view of the \hst\ observations to directly register the images, hence our use
of \gaia\ DR2 as an intermediary.} \citep{GaiaDR2}. For \chandra, the image
transformation is determined using \wcsmatch\ and then applied to the combined image
using \wcsupdate, both part of \ciao. The transformation is determined by initially
matching sources within a 2$''$ radius, and then iteratively updating the astrometric
solution to keep only those that match within a radius of 0.5$''$ once the transformation
is applied; this results in 5 matches with the \gaia\ catalogue (note that none of these are
the ULX in question), and leaves a residual uncertainty on the astrometric solution of
0.26$''$ (1$\sigma$ confidence). The position of ULX3 in the updated \chandra\ image is
RA = 21$^{h}$36$^{m}$22.74$^{s}$, DEC = $-$54{\deg}32$'$33.8$''$. This has a
statistical precision of 0.09$''$, giving a total 3$\sigma$ uncertainty on the position of
ULX3 of 0.83$''$ (combining the statistical and astrometric uncertainties in quadrature).
Although we do not repeat this analysis for the \xmm\ data, we note that there are no
other X-ray sources detected by \chandra\ within 50$''$, and that the raw \xmm\ centroid
position from epoch XN1 is in excellent agreement with the \chandra\ position given
above. For \hst, we use the `Match to star positions' tool in the STARLINK/GAIA software
to update the astrometric solutions of the drizzled images retrieved from the
\textit{Hubble} Legacy Archive. We find 7-12 good matches in each image. The resulting
astrometric uncertainties are smaller than 0.05$''$ (1$\sigma$ confidence) for all images
and thus provide a negligible contribution to the localization uncertainty of the ULX.

\begin{figure}
\includegraphics[width=0.475\textwidth]{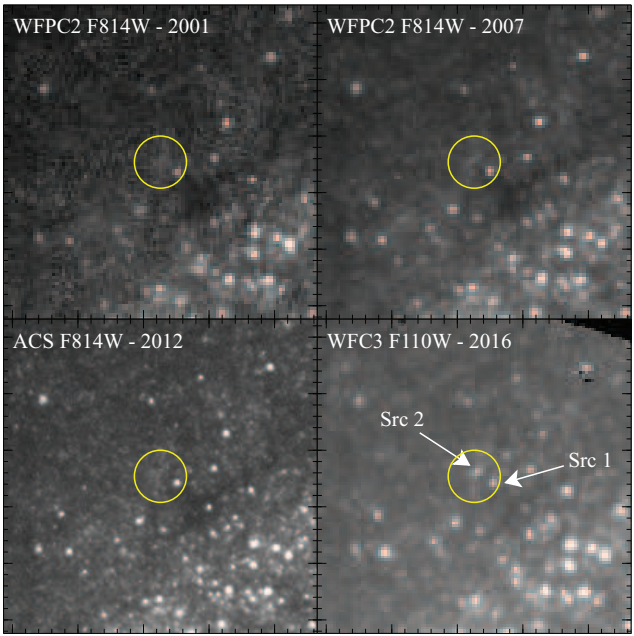}
\caption{Drizzled \hst\ images in the F814W and F110W filters. Each panel shows a
$10'' \times 10''$ region around the location of the ULX. North is up, East to the left.
The yellow circles indicate the 0.83$''$ radius, 3$\sigma$ confidence localization of the
ULX.}
\label{fig_hst}
\end{figure}

We detect two potential counterparts in the \hst\ images. One is located on the edge of
the error circle and clearly detected in all images. The second is located closer to the
center of the error circle but only clearly visible in the most recent, WFC3 near-IR
images. Both sources are indicated in the bottom-right panel in Fig.~\ref{fig_hst}.

We obtain PSF photometry of the two candidate counterparts with {\sc Dolphot}
\citep{Dolphot}. Following the recommendations in the manual, we use the c0m and
c1m (WFPC2) and flt (ACS, WFC3) images for the photometric analysis. We use the 
recommended parameter settings for each instrument, with fitsky = 2. The computed
instrumental Vega magnitudes for the sources indicated in Fig.~\ref{fig_hst} are listed
in Table \ref{tab_hst}. Source 1 is clearly detected in all epochs and filters; source 2 is
(sometimes marginally) detected in all but the first epochs. We determine the limiting
magnitude for the 2001 WFPC2 observations using the {\it fakestars} routine in
{\sc Dolphot}, by simulating and retrieving 2000 stars with instrumental magnitudes in
the range of 22-27 around the position of the ULX. We adopt the magnitude where
90\% of simulated stars are still detected with S/N $\geq 5$ as our limiting magnitude.
Source 1 does not show variability in the F814W band, the only filter for which multiple
epochs are available, while source 2 appears to be somewhat variable; unfortunately
all F814W observations were obtained prior to the last \swift\ observation (in 2014)
where the ULX was undetected, so we cannot investigate whether a change in the
optical properties of the source accompanied the increase in the X-ray luminosity.
We also subtract an appropriately scaled version of the ACS/F625W image from the
ACS/F658N image to search for excess H$\alpha$ emission associated with the ULX.
However, we do not detect point-like excess H$\alpha$ emission associated with either
of the candidate counterparts in the continuum-subtracted image. The ULX does lie in
a region with faint extended H$\alpha$ emission, but this appears to be part of a larger
structure not necessarily associated with the ULX itself.

The Galactic foreground extinction in the direction of NGC~7090 is low ($A_V = 0.063$,
\citealt{Schlafly11}), but given that it is an edge-on spiral galaxy and that the ULX
appears to be located on one of the dust lanes, there must be local extinction as well.
Assuming $N_\mathrm{H} = 2.21 \times 10^{21} A_V$\,\pcmsq\ \citep{Guver09} and
$N_\mathrm{H} \approx 7.9 \times 10^{21}$\,\pcmsq, we find $A_V \approx 3.6$. As
part of this hydrogen column may be intrinsic to the X-ray source we consider this value
as an upper limit to the extinction of the optical emission. With $A_V$ in the range of
$1-3.6$ and $m-M = 29.9$, source 1 has absolute magnitudes and colors roughly
consistent with a blue or yellow supergiant (potentially similar to NGC\,7793 P13;
\citealt{Motch14nat}). Similar to other ULXs with blue counterparts, the emission could
also be dominated by an irradiated accretion disc \citep[e.g.,][]{Roberts11, Tao12},
although we note again that all of the optical observations appear to have been taken
while ULX3 was in its lower luminosity state ($L_{\rm{X}} \sim 10^{38}$\,\ergps, as far
as it is possible to tell from the limited temporal coverage; see Figure \ref{fig_longlc}).
Compared to the range of absolute magnitudes observed in Galactic low-mass X-ray
binaries in outburst (which often reach comparable X-ray luminosities to \ulx3\ in its
low-luminosity state), this candidate optical counterpart is on the bright end of the
distribution \citep{Vanparadijs94}, but again we cannot exclude the possibility that the
donor is a lower mass star and the optical emission is dominated by an irradiated
accretion disc.

Source 2 has absolute magnitudes and colors roughly consistent with those of red
supergiants (potentially similar to NGC\,300 ULX1; \citealt{Heida19}). However, as this
source displays variability in the F814W band, colors calculated from observations
taken several years apart are obviously unreliable; multi-band photometry taken at a
single epoch, as well as a better determination of the local reddening, is necessary to
determine the nature of this candidate counterpart. Detailed optical spectroscopy of
these counterparts is not currently plausible, but may be possible with the next
generation of 30m-class telescopes.

\section{Discussion \& Conclusions}
\label{sec_dis}

\ulx3\ -- aka \namcsc\ and \namxmm\ -- is a newly discovered ULX in the galaxy \ngc,
with a peak luminosity of $L_{\rm{X,peak}} \sim 6 \times 10^{39}$\,\ergps. This is the
latest member of the growing population of transient ULXs (\eg\ \citealt{Soria12,
Middleton12, Middleton13nat, Pintore18, Pintore20, Earnshaw19, Earnshaw20,
Brightman20m51}). Remarkably, ULX3 is the third such source in the galaxy \ngc\
alone (\eg\ \citealt{Liu19, Song20}).

Although we refer to ULX3 as a transient ULX, as it has only recently been seen to
exhibit luminosities at the ULX level, it is not necessarily an X-ray transient in the more
traditional sense. Most X-ray binaries (XRBs) in our own Galaxy are transient LMXBs,
which spend the majority of the time in quiescence ($L_{\rm{X}} \sim
10^{30-34}$\,\ergps; \eg\ \citealt{Homan13, Reynolds14quies}), interspersed by
transient outbursts of activity reaching much higher luminosities (which are widely
expected to be related to the hydrogen ionisation instability; \citealt{Lasota01rev}).
Although a rare occurence, some of these sources can reach peak luminosities similar
to ULX3 (\eg\ \citealt{Middleton12}). However, ULX3 appeared to have a relatively
stable luminosity of $\sim$10$^{38}$\,\ergps\ prior to its recent transition into the ULX
regime. This potentially causes ULX3 to stand out from classic LMXBs, as such
luminosities would very much be in the outburst regime for such sources. If either of the
candidate optical counterparts seen in the \hst\ observations are dominated by the
donor star, ULX3 could be accreting from a supergiant companion. However, the large
increase in luminosity seen recently means ULX3 also stands out from classic wind-fed
HMXBs, which tend to be (relatively) persistent. Some kind of Be/XRB-like phenomenon
might be possible; in addition to the regular outbursts that occur when the compact
object passes through the decretion disc that surrounds its stellar companion, these
sources occasionally exhibit rare `type II' outbursts which can reach super-Eddington
luminosities and appear to be unrelated to the orbital dynamics of the system (see
\citealt{Reig11} for a review on Be/XRB systems). However, here too the apparently
stable luminosity seen from ULX3 would be abnormally high. Unfortunately, given the
sparse coverage, we can only place very loose constraints on the timescale over which
ULX3 evolved into the ULX regime; given that the source was not detected in any of the
individual \swift\ observations prior to our recent observing campaign, this must have
occurred sometime between July 2014 and October 2019, a window of $\sim$5 years.
However, this period of activity seems to have lasted $>$7 months, as ULX3 has been
almost persistently detected by \swift\ over our 2019--20 monitoring campaign; this
would also be abnormally long for a type II outburst from a Be/XRB-like system.

The nature of the accretor in ULX3 is not clear from the current data. On the one hand,
the spectral data are consistent with being dominated by emission from a standard thin
disc with a constant inner radius for all observing epochs to date (implying in turn that
the data are consistent with $L \propto T_{\rm{in}}^{4}$, although we stress again that
the uncertainties are large for the low-flux data). This may suggest the presence of a
black hole accretor. Taking these results at face value, we estimate a minimum value
for the inner radius from the normalisation of the \diskbb\ model for epoch
XN1,\footnote{We also investigated a joint fit to all of the \xmm, \nustar\ and \chandra\
data (new and archival) with the \diskbb\ normalisation linked across all epochs, but
found that the results were no different to fitting epoch XN1 alone, as this dominates
the total S/N.} as this is given by $[R_{\rm{in}}/(D \xi f_{\rm{col}}^{2})]^2 \cos(i)$. Here,
$R_{\rm{in}}$ and $D$ are in units of km and 10\,kpc, respectively, while $f_{\rm{col}}$
and $\xi$ are corrections that account for the complex atmospheric physics in the disc
and the fact that the peak temperature actually arises at a radius slightly larger than
\rin, respectively. The product $\xi f_{\rm{col}}^{2}$ is generally taken to be $\sim$1.2
for a standard thin disc (\citealt{Shimura95, Kubota98}). Taking $\cos(i) = 1$ as a
limiting case, we find $R_{\rm{in}} \gtrsim 55$\,km. 

This is significantly larger than the standard neutron star radius ($R_{\rm{NS}} \sim
13$\,km; \citealt{Riley19, Miller19}) and assuming the disc reaches the innermost stable
circular orbit of the accretor, would imply the presence of a black hole with a minimum
mass of $M_{\rm{BH}} > 6-30$\,\msun\ (for spin parameters $0 \leq a^* \leq 0.998$, \ie
\rin\ = 1--6\,\rg, where \rg\ = $GM_{\rm{BH}}/c^{2}$). If we instead assume an
intermediate inclination ($\cos(i) = 0.5$) and an intermediate spin (such that \rin\ =
3\,\rg), then this would imply \rin\ $\sim$ 75\,km and in turn \mbh\ $\sim$ 20\,\msun. In
addition, the thin nature of the best-fit accretion disc model would imply that the
luminosity never exceeds the source's Eddington limit, if also taken at face value. 
Although we have previously assumed isotropic emission, if the emission is truly
dominated by a thin accretion disc then the inclination of the disc has to be accounted
for when estimating the peak luminosity, and we find $L_{\rm{X,peak}} \sim 3 \times
10^{39}/\cos(i)$\,\ergps. For the average inclination expected for a randomly orientated
disc, \ie $\cos(i) = 0.5$, we return to $L_{\rm{X,peak}} \sim 6 \times 10^{39}$\,\ergps,
which would in turn imply a lower limit to the BH mass of $M_{\rm{BH}} \gtrsim
40$\,\msun. This mass would be consistent with the \diskbb\ normalisation for $a^* =
0.998$. Interestingly, the disc temperature expected for a rapidly rotating 40\,\msun\
black hole accreting at close to its Eddington limit is $\sim$2\,keV (\eg\
\citealt{Makishima00}), similar to that observed from epoch XN1. Indeed, if we fit epoch
XN1 with \kerrbb\footnote{Note that we use the updated version described in
\cite{Parker19}.}, a fully relativistic thin disc model (\citealt{kerrbb}), instead of the
simpler \diskbb\ model, fixing the spin to the maximal value, the inclination to 60\deg
and assuming a standard colour correction factor of $f_{\rm{col}} = 1.7$
(\citealt{Shimura95, Davis19}), the best-fit black hole mass is $\sim$37\,\msun.

\begin{figure}
\begin{center}
\hspace*{-0.35cm}
\rotatebox{0}{
{\includegraphics[width=235pt]{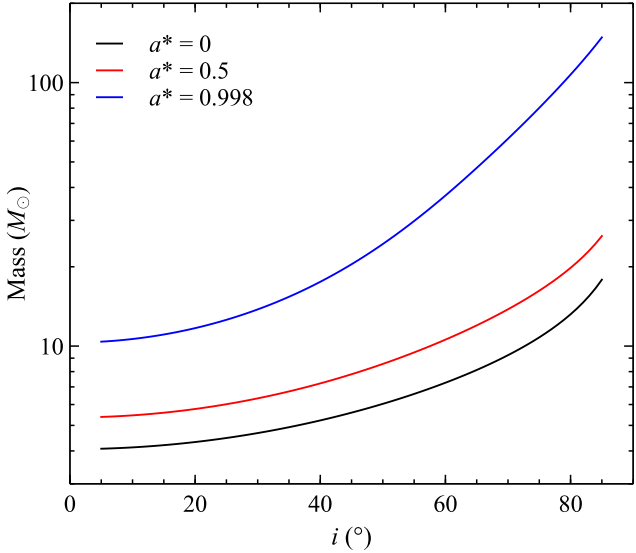}}
}
\end{center}
\vspace*{-0.3cm}
\caption{The dependence of the best-fit mass from the \kerrbb\ model (\citealt{kerrbb})
on the inclination of the disc, assuming $f_{\rm{col}} = 1.7$, for three different spin
parameters: non-rotating ($a^* = 0$), moderately rotating ($a^* = 0.5$) and maximally rotating ($a^* = 0.998$).
}
\label{fig_mass}
\end{figure}

Although such a black hole would be larger than any seen in an X-ray binary in our own
galaxy (\eg\ \citealt{Orosz03}), similar mass BHs are known to exist as they are now
fairly regularly being seen in BH--BH mergers by LIGO (\citealt{GWTC1}). The formation
of such a black hole via standard stellar evolution may require a low metallicity (\eg\
\citealt{Zampieri09, Belczynski10}). Although there is not much information regarding the
metallicity of NGC\,7090 available in the literature, we note that the majority of oxygen
abundance estimates compiled by \cite{deVis19} would imply an abundance of
$A_{\rm{O}}$/solar $\sim$ 0.5 (or alternatively $12 + \log[\rm{O/H}] \sim 8.4$). Smaller
masses could still formally be permitted for other combinations of spin and inclination;
we show the dependence of the best-fit mass on the disc inclination for three spin
parameters (non-rotating, moderately rotating and maximally rotating) based on the
\kerrbb\ model in Figure \ref{fig_mass}. However, this would push the peak luminosity
into the super-Eddington regime, for which the \kerrbb\ model is not formally valid.
Furthermore, the colour correction factor used here may not be valid for accretion
$\sim$at/above the Eddington limit; higher values may be expected instead (\eg\
\citealt{Watari03, Kawaguchi03}), in which case the mass estimates assuming
$f_{\rm{col}} = 1.7$ would be underestimated by a factor of ($f_{\rm{col}}/1.7)^{2}$.
\cite{Watari03} suggest that $f_{\rm{col}} \sim 3$, in which case the best-fit mass curves
shown in Figure \ref{fig_mass} would systematically shift upwards by a factor of $\sim$3.

On the other hand, the long-term flux distribution (although still fairly sparsely sampled
in terms of sensitive observations) is consistent with ULX3 having a bi-modal flux
distribution. This sort of distribution is broadly expected for neutron stars undergoing
transitions to/from the propeller regime (\citealt{Tsygankov16}), as the infalling material
cannot pass through the magnetosphere (\rmag) of the neutron star in the propeller
regime, and so at some transition point the accretion rate (and thus the observed
luminosity) drops precipitously. It may well be that flux distributions similar to that seen
here are tentative evidence for a magnetic neutron star accretor. Many of the known
ULX pulsars exhibit low-flux states (\eg\ \citealt{Motch14nat, Walton15}), and evidence
for similar behaviour is now being seen in a growing number of sources among the
broader ULX population (\citealt{Earnshaw18, Song20}). These events are potentially
related to the propeller regime, requiring a magnetized neutron star accretor, although
other possibilities certainly remain possible. Most notably, obscuration of the inner
accretion flow by its outer regions/winds has also been invoked to explain
high-amplitude variability events in some cases (\eg\ \citealt{Vasilopoulos19}),
particularly where long-timescale ($\gtrsim$10s of days) super-orbital X-ray periods with
large variability amplitudes are present (\citealt{Brightman19m82, Brightman20m51,
Vasilopoulos20}; see also \citealt{Middleton15, Middleton18}). While such an
explanation would potentially need ULX3 to have an \textit{extremely} long X-ray period,
given the apparent stability of the low-flux state across many years prior to our recent
observations, it is also interesting to note that the long-timescale X-ray periods in ULXs
are themselves most robustly seen in the known ULX pulsar systems (\eg\
\citealt{Walton16period, Hu17, Fuerst18, Brightman19m82, Brightman20m51,
Vasilopoulos20}). Although we do not have any significant detection of X-ray pulsations
from ULX3, the current limits are only mildly constraining, and the pulsations are seen
to be transient in a number of the known ULX pulsars (\eg\ \citealt{Israel17,
Sathyaprakash19, Bachetti20}). Furthermore, as discussed earlier, the lack of spectral
complexity similar to that seen in other known ULX pulsars could easily be purely due
to a combination of relatively high absorption and low S/N at high energies. The latter in
particular could easily prevent us from significantly detecting the emission from any
central accretion columns. 

In the magnetized neutron star scenario, the inner radius of the disc is set by the
magnetospheric radius, rather than the radius of the neutron star itself. Standard
accretion theory for magnetic neutron stars (\ie\ assuming a dipolar field geometry and a
thin accretion disc, the latter of which may not be formally appropriate here) implies that
the magnetospheric radius is given by $R_{\rm{M}} = (3.9 \times 10^{8}) L^{-2/7}_{37}
B^{4/7}_{12}$ for a 1.4\,\msun\ neutron star with a radius of 13\,km, where $L_{37}$ is
the X-ray luminosity in units of $10^{37}$\,\ergps\ and $B_{12}$ is the magnetic field
strength in units of $10^{12}$\,G (\citealt{Lamb73, Cui97}). The inner radius of 76\,km
estimated previously would therefore imply a rather weak field of $\sim$2 $\times$
10$^{10}$\,G. Typical field strengths for X-ray pulsars in our own Galaxy are
$\sim$10$^{12}$\,G (\citealt{Caballero12rev}), although it has previously been
suggested that ULX pulsars specifically may have weak fields (\eg\ \citealt{Kluzniak15,
King16}). It is also worth noting, however, that stronger fields could still be
accommodated for an equivalent $R_{\rm{M}}$ if the field geometry is higher order than
the standard dipole assumed above, which is a possibility that has also been suggested
for ULX pulsars (\eg\ \citealt{Israel17}).

Ultimately, however, further observations that can provide improved constraints on the
timing properties and/or the evolution of the accretion flow will be required to reveal the
nature of ULX3.

%

\section*{ACKNOWLEDGEMENTS}

The authors would like to thank the reviewer for their positive feedback, which helped
to improve the final version of the manuscript.
DJW and MJM acknowledge support from the Science and Technology Facilities Council
(STFC) in the form of Ernest Rutherford Fellowships.
PAE acknowledges UK Space Agency (UKSA) support.
This research has made use of data obtained with \nustar, a project led by Caltech,
funded by NASA and managed by the NASA Jet Propulsion Laboratory (JPL), and has
utilized the \nustardas\ software package, jointly developed by the Space Science
Data Centre (SSDC; Italy) and Caltech (USA).
This research has also made use of data obtained with \xmm, an ESA science mission
with instruments and contributions directly funded by ESA Member States, as well as
public data from the \swift\ data archive. 
Finally, this work has also made use of data obtained from the \chandra\ Source
Catalog, provided by the \chandra\ X-ray Center (CXC) as part of the \chandra\ Data
Archive. 


\section*{Data Availability}

All of the data underlying this article are either already publicly available from ESA's
\xmm\ Science Archive (https://www.cosmos.esa.int/web/xmm-newton/xsa), NASA's
HEASARC archive (https://heasarc.gsfc.nasa.gov/) and NASA's \chandra\ Data
Archive (https://cxc.harvard.edu/cda/), or will be from May 2021.

\bibliographystyle{../../mnras}

\bibliography{../../references}

\begin{thebibliography}{133}
\expandafter\ifx\csname natexlab\endcsname\relax\def\natexlab#1{#1}\fi

\bibitem[Abbott et~al.(2019)Abbott, Abbott, Abbott et~al.]{GWTC1}
Abbott B., Abbott R., Abbott T., et~al., 2019, {Phys.Rev.X}, 9, 3, 031040

\bibitem[{Abramowicz} et~al.(1988){Abramowicz}, {Czerny}, {Lasota} \&
  {Szuszkiewicz}]{Abram88}
{Abramowicz} M.~A., {Czerny} B., {Lasota} J.~P., {Szuszkiewicz} E., 1988, \apj,
  332, 646

\bibitem[{Arnaud}(1996)]{xspec}
{Arnaud} K.~A., 1996, in { Astronomical Data Analysis Software and Systems
  V\/}, edited by {G.~H.~Jacoby \& J.~Barnes}, vol. 101 of { Astron. Soc. Pac.
  Conference Series, Astron. Soc. Pac., San Francisco\/}, ~17

\bibitem[{Ba{\~n}ados} et~al.(2018){Ba{\~n}ados}, {Venemans}, {Mazzucchelli}
  et~al.]{Banados18nat}
{Ba{\~n}ados} E., {Venemans} B.~P., {Mazzucchelli} C., et~al., 2018, \nat, 553,
  473

\bibitem[Bachetti(2018)]{bachettiHENDRICSHighENergy2018}
Bachetti M., 2018, Astrophysics Source Code Library,  ascl:1805.019

\bibitem[{Bachetti} et~al.(2014){Bachetti}, {Harrison}, {Walton}
  et~al.]{Bachetti14nat}
{Bachetti} M., {Harrison} F.~A., {Walton} D.~J., et~al., 2014, \nat, 514, 202

\bibitem[{Bachetti} et~al.(2020){Bachetti}, {Maccarone}, {Brightman}
  et~al.]{Bachetti20}
{Bachetti} M., {Maccarone} T.~J., {Brightman} M., et~al., 2020, \apj, 891, 1,
  44

\bibitem[{Bachetti} et~al.(2013){Bachetti}, {Rana}, {Walton}
  et~al.]{Bachetti13}
{Bachetti} M., {Rana} V., {Walton} D.~J., et~al., 2013, \apj, 778, 163

\bibitem[{Belczynski} et~al.(2010){Belczynski}, {Bulik}, {Fryer}
  et~al.]{Belczynski10}
{Belczynski} K., {Bulik} T., {Fryer} C.~L., et~al., 2010, \apj, 714, 1217

\bibitem[{Brightman} et~al.(2020){Brightman}, {Earnshaw}, {F{\"u}rst}
  et~al.]{Brightman20m51}
{Brightman} M., {Earnshaw} H., {F{\"u}rst} F., et~al., 2020, \apj, 895, 2, 127

\bibitem[{Brightman} et~al.(2016){Brightman}, {Harrison}, {Walton}
  et~al.]{Brightman16m82a}
{Brightman} M., {Harrison} F., {Walton} D.~J., et~al., 2016, \apj, 816, 60

\bibitem[{Brightman} et~al.(2019){Brightman}, {Harrison}, {Bachetti}
  et~al.]{Brightman19m82}
{Brightman} M., {Harrison} F.~A., {Bachetti} M., et~al., 2019, \apj, 873, 2,
  115

\bibitem[{Brightman} et~al.(2018){Brightman}, {Harrison}, {F{\"u}rst}
  et~al.]{Brightman18}
{Brightman} M., {Harrison} F.~A., {F{\"u}rst} F., et~al., 2018, Nature
  Astronomy, 2, 312

\bibitem[Buccheri et~al.(1983)Buccheri, Bennett, Bignami et~al.]{Buccheri83}
Buccheri R., Bennett K., Bignami G.~F., et~al., 1983, A\&A, 128, 245

\bibitem[{Caballero} \& {Wilms}(2012)]{Caballero12rev}
{Caballero} I., {Wilms} J., 2012, \memsai, 83, 230

\bibitem[{Carpano} et~al.(2018){Carpano}, {Haberl}, {Maitra} \&
  {Vasilopoulos}]{Carpano18}
{Carpano} S., {Haberl} F., {Maitra} C., {Vasilopoulos} G., 2018, \mnras, 476,
  L45

\bibitem[{Cash}(1979)]{cstat}
{Cash} W., 1979, \apj, 228, 939

\bibitem[{Chandra} et~al.(2020){Chandra}, {Roy}, {Agrawal} \&
  {Choudhury}]{Chandra20}
{Chandra} A.~D., {Roy} J., {Agrawal} P.~C., {Choudhury} M., 2020, \mnras, 495,
  3, 2664

\bibitem[{Cui}(1997)]{Cui97}
{Cui} W., 1997, \apjl, 482, L163

\bibitem[{Davis} \& {El-Abd}(2019)]{Davis19}
{Davis} S.~W., {El-Abd} S., 2019, \apj, 874, 23

\bibitem[{De Vis} et~al.(2019){De Vis}, {Jones}, {Viaene} et~al.]{deVis19}
{De Vis} P., {Jones} A., {Viaene} S., et~al., 2019, \aap, 623, A5

\bibitem[{Dolphin}(2000)]{Dolphot}
{Dolphin} A.~E., 2000, \pasp, 112, 1383

\bibitem[{Earnshaw} et~al.(2019{\natexlab{a}}){Earnshaw}, {Grefenstette},
  {Brightman} et~al.]{Earnshaw19}
{Earnshaw} H.~P., {Grefenstette} B.~W., {Brightman} M., et~al.,
  2019{\natexlab{a}}, \apj, 881, 1, 38

\bibitem[{Earnshaw} et~al.(2020){Earnshaw}, {Heida}, {Brightman}
  et~al.]{Earnshaw20}
{Earnshaw} H.~P., {Heida} M., {Brightman} M., et~al., 2020, \apj, 891, 2, 153

\bibitem[{Earnshaw} et~al.(2019{\natexlab{b}}){Earnshaw}, {Roberts},
  {Middleton}, {Walton} \& {Mateos}]{EarnshawULXcat}
{Earnshaw} H.~P., {Roberts} T.~P., {Middleton} M.~J., {Walton} D.~J., {Mateos}
  S., 2019{\natexlab{b}}, \mnras, 483, 4, 5554

\bibitem[{Earnshaw} et~al.(2018){Earnshaw}, {Roberts} \&
  {Sathyaprakash}]{Earnshaw18}
{Earnshaw} H.~P., {Roberts} T.~P., {Sathyaprakash} R., 2018, \mnras, 476, 4272

\bibitem[{El Mellah} et~al.(2019){El Mellah}, {Sundqvist} \&
  {Keppens}]{ElMellah19}
{El Mellah} I., {Sundqvist} J.~O., {Keppens} R., 2019, \aap, 622, L3

\bibitem[{Evans} et~al.(2020){Evans}, {Primini}, {Miller} et~al.]{CSC2temp}
{Evans} I.~N., {Primini} F.~A., {Miller} J.~B., et~al., 2020, in { American
  Astronomical Society Meeting Abstracts\/}, American Astronomical Society
  Meeting Abstracts,  154.05

\bibitem[{Evans} et~al.(2009){Evans}, {Beardmore}, {Page} et~al.]{Evans09}
{Evans} P.~A., {Beardmore} A.~P., {Page} K.~L., et~al., 2009, \mnras, 397, 1177

\bibitem[{F{\"u}rst} et~al.(2016){F{\"u}rst}, {Walton}, {Harrison}
  et~al.]{Fuerst16p13}
{F{\"u}rst} F., {Walton} D.~J., {Harrison} F.~A., et~al., 2016, \apjl, 831, L14

\bibitem[{F{\"u}rst} et~al.(2018){F{\"u}rst}, {Walton}, {Heida}
  et~al.]{Fuerst18}
{F{\"u}rst} F., {Walton} D.~J., {Heida} M., et~al., 2018, \aap, 616, A186

\bibitem[{F{\"u}rst} et~al.(2017){F{\"u}rst}, {Walton}, {Stern}
  et~al.]{Fuerst17ngc5907}
{F{\"u}rst} F., {Walton} D.~J., {Stern} D., et~al., 2017, \apj, 834, 77

\bibitem[{Gaia Collaboration} et~al.(2018){Gaia Collaboration}, {Brown},
  {Vallenari} et~al.]{GaiaDR2}
{Gaia Collaboration}, {Brown} A.~G.~A., {Vallenari} A., et~al., 2018, \aap,
  616, A1

\bibitem[{Garmire} et~al.(2003){Garmire}, {Bautz}, {Ford}, {Nousek} \&
  {Ricker}]{CHANDRA_ACIS}
{Garmire} G.~P., {Bautz} M.~W., {Ford} P.~G., {Nousek} J.~A., {Ricker} Jr.
  G.~R., 2003, in { SPIE Conference Series\/}, edited by J.~E. {Truemper},
  H.~D. {Tananbaum}, vol. 4851 of { SPIE Conference Series\/},  28--44

\bibitem[{Gehrels} et~al.(2004){Gehrels}, {Chincarini}, {Giommi} et~al.]{SWIFT}
{Gehrels} N., {Chincarini} G., {Giommi} P., et~al., 2004, \apj, 611, 1005

\bibitem[{Gladstone} et~al.(2009){Gladstone}, {Roberts} \& {Done}]{Gladstone09}
{Gladstone} J.~C., {Roberts} T.~P., {Done} C., 2009, \mnras, 397, 1836

\bibitem[{G{\"u}ver} \& {{\"O}zel}(2009)]{Guver09}
{G{\"u}ver} T., {{\"O}zel} F., 2009, \mnras, 400, 4, 2050

\bibitem[{Harrison} et~al.(2013){Harrison}, {Craig}, {Christensen}
  et~al.]{NUSTAR}
{Harrison} F.~A., {Craig} W.~W., {Christensen} F.~E., et~al., 2013, \apj, 770,
  103

\bibitem[{Heida} et~al.(2016){Heida}, {Jonker}, {Torres} et~al.]{Heida16}
{Heida} M., {Jonker} P.~G., {Torres} M.~A.~P., et~al., 2016, \mnras, 459, 771

\bibitem[{Heida} et~al.(2019){Heida}, {Lau}, {Davies} et~al.]{Heida19}
{Heida} M., {Lau} R.~M., {Davies} B., et~al., 2019, \apjl, 883, 2, L34

\bibitem[{Heida} et~al.(2015){Heida}, {Torres}, {Jonker} et~al.]{Heida15}
{Heida} M., {Torres} M.~A.~P., {Jonker} P.~G., et~al., 2015, \mnras, 453, 3511

\bibitem[{HI4PI Collaboration} et~al.(2016){HI4PI Collaboration}, {Ben Bekhti},
  {Fl{\"o}er} et~al.]{NH2016}
{HI4PI Collaboration}, {Ben Bekhti} N., {Fl{\"o}er} L., et~al., 2016, \aap,
  594, A116

\bibitem[{Homan} et~al.(2013){Homan}, {Fridriksson}, {Jonker} et~al.]{Homan13}
{Homan} J., {Fridriksson} J.~K., {Jonker} P.~G., et~al., 2013, \apj, 775, 1, 9

\bibitem[{Hu} et~al.(2017){Hu}, {Li}, {Kong}, {Ng} \& {Chun-Che Lin}]{Hu17}
{Hu} C.-P., {Li} K.~L., {Kong} A.~K.~H., {Ng} C.-Y., {Chun-Che Lin} L., 2017,
  \apjl, 835, L9

\bibitem[{Israel} et~al.(2017{\natexlab{a}}){Israel}, {Belfiore}, {Stella}
  et~al.]{Israel17}
{Israel} G.~L., {Belfiore} A., {Stella} L., et~al., 2017{\natexlab{a}},
  Science, 355, 817

\bibitem[{Israel} et~al.(2017{\natexlab{b}}){Israel}, {Papitto}, {Esposito}
  et~al.]{Israel17p13}
{Israel} G.~L., {Papitto} A., {Esposito} P., et~al., 2017{\natexlab{b}},
  \mnras, 466, L48

\bibitem[{Jansen} et~al.(2001){Jansen}, {Lumb}, {Altieri} et~al.]{XMM}
{Jansen} F., {Lumb} D., {Altieri} B., et~al., 2001, \aap, 365, L1

\bibitem[{Kaaret} et~al.(2017){Kaaret}, {Feng} \& {Roberts}]{Kaaret17rev}
{Kaaret} P., {Feng} H., {Roberts} T.~P., 2017, \araa, 55, 303

\bibitem[{Karachentsev} et~al.(2018){Karachentsev}, {Kaisina} \&
  {Makarov}]{Karachentsev18}
{Karachentsev} I.~D., {Kaisina} E.~I., {Makarov} D.~I., 2018, \mnras, 479, 3,
  4136

\bibitem[{Kawaguchi}(2003)]{Kawaguchi03}
{Kawaguchi} T., 2003, \apj, 593, 1, 69

\bibitem[{King} \& {Lasota}(2016)]{King16}
{King} A., {Lasota} J.-P., 2016, \mnras, 458, L10

\bibitem[{King} et~al.(2001){King}, {Davies}, {Ward}, {Fabbiano} \&
  {Elvis}]{King01}
{King} A.~R., {Davies} M.~B., {Ward} M.~J., {Fabbiano} G., {Elvis} M., 2001,
  \apjl, 552, L109

\bibitem[{King} \& {Pounds}(2003)]{King03}
{King} A.~R., {Pounds} K.~A., 2003, \mnras, 345, 657

\bibitem[{Klu{\'z}niak} \& {Lasota}(2015)]{Kluzniak15}
{Klu{\'z}niak} W., {Lasota} J.-P., 2015, \mnras, 448, L43

\bibitem[{Kosec} et~al.(2018){Kosec}, {Pinto}, {Walton} et~al.]{Kosec18}
{Kosec} P., {Pinto} C., {Walton} D.~J., et~al., 2018, \mnras, 479, 3978

\bibitem[{Kubota} et~al.(1998){Kubota}, {Tanaka}, {Makishima} et~al.]{Kubota98}
{Kubota} A., {Tanaka} Y., {Makishima} K., et~al., 1998, \pasj, 50, 667

\bibitem[{Lamb} et~al.(1973){Lamb}, {Pethick} \& {Pines}]{Lamb73}
{Lamb} F.~K., {Pethick} C.~J., {Pines} D., 1973, \apj, 184, 271

\bibitem[{Lasota}(2001)]{Lasota01rev}
{Lasota} J.-P., 2001, NewAR, 45, 449

\bibitem[Leahy et~al.(1983)Leahy, Darbro, Elsner et~al.]{Leahy83b}
Leahy D.~A., Darbro W., Elsner R.~F., et~al., 1983, ApJ, 266, 160

\bibitem[{Lehmer} et~al.(2019){Lehmer}, {Eufrasio}, {Tzanavaris}
  et~al.]{Lehmer19}
{Lehmer} B.~D., {Eufrasio} R.~T., {Tzanavaris} P., et~al., 2019, \apjs, 243, 1,
  3

\bibitem[{Li} et~al.(2005){Li}, {Zimmerman}, {Narayan} \& {McClintock}]{kerrbb}
{Li} L.-X., {Zimmerman} E.~R., {Narayan} R., {McClintock} J.~E., 2005, \apjs,
  157, 335

\bibitem[{Liu} et~al.(2019){Liu}, {O'Brien}, {Osborne}, {Evans} \&
  {Page}]{Liu19}
{Liu} Z., {O'Brien} P.~T., {Osborne} J.~P., {Evans} P.~A., {Page} K.~L., 2019,
  \mnras, 486, 4, 5709

\bibitem[{Madsen} et~al.(2020){Madsen}, {Grefenstette}, {Pike}
  et~al.]{NuSTARmli}
{Madsen} K.~K., {Grefenstette} B.~W., {Pike} S., et~al., 2020, arXiv e-prints,
  arXiv:2005.00569

\bibitem[{Madsen} et~al.(2015){Madsen}, {Harrison}, {Markwardt}
  et~al.]{NUSTARcal}
{Madsen} K.~K., {Harrison} F.~A., {Markwardt} C.~B., et~al., 2015, \apjs, 220,
  8

\bibitem[{Makishima} et~al.(2000){Makishima}, {Kubota}, {Mizuno}
  et~al.]{Makishima00}
{Makishima} K., {Kubota} A., {Mizuno} T., et~al., 2000, \apj, 535, 632

\bibitem[{Mart{\'\i}nez-N{\'u}{\~n}ez}
  et~al.(2017){Mart{\'\i}nez-N{\'u}{\~n}ez}, {Kretschmar}, {Bozzo}
  et~al.]{MartinezNunez17rev}
{Mart{\'\i}nez-N{\'u}{\~n}ez} S., {Kretschmar} P., {Bozzo} E., et~al., 2017,
  \ssr, 212, 1-2, 59

\bibitem[{Middleton} et~al.(2018){Middleton}, {Fragile}, {Bachetti}
  et~al.]{Middleton18}
{Middleton} M.~J., {Fragile} P.~C., {Bachetti} M., et~al., 2018, \mnras, 475,
  1, 154

\bibitem[{Middleton} et~al.(2015){Middleton}, {Heil}, {Pintore}, {Walton} \&
  {Roberts}]{Middleton15}
{Middleton} M.~J., {Heil} L., {Pintore} F., {Walton} D.~J., {Roberts} T.~P.,
  2015, \mnras, 447, 3243

\bibitem[{Middleton} et~al.(2013){Middleton}, {Miller-Jones}, {Markoff}
  et~al.]{Middleton13nat}
{Middleton} M.~J., {Miller-Jones} J.~C.~A., {Markoff} S., et~al., 2013, \nat,
  493, 187

\bibitem[{Middleton} et~al.(2012){Middleton}, {Sutton}, {Roberts}, {Jackson} \&
  {Done}]{Middleton12}
{Middleton} M.~J., {Sutton} A.~D., {Roberts} T.~P., {Jackson} F.~E., {Done} C.,
  2012, \mnras, 420, 2969

\bibitem[{Miller} et~al.(2019){Miller}, {Lamb}, {Dittmann} et~al.]{Miller19}
{Miller} M.~C., {Lamb} F.~K., {Dittmann} A.~J., et~al., 2019, \apjl, 887, 1,
  L24

\bibitem[{Mineo} et~al.(2012){Mineo}, {Gilfanov} \& {Sunyaev}]{Mineo12}
{Mineo} S., {Gilfanov} M., {Sunyaev} R., 2012, \mnras, 419, 3, 2095

\bibitem[{Mineshige} et~al.(1994){Mineshige}, {Hirano}, {Kitamoto}, {Yamada} \&
  {Fukue}]{diskpbb}
{Mineshige} S., {Hirano} A., {Kitamoto} S., {Yamada} T.~T., {Fukue} J., 1994,
  \apj, 426, 308

\bibitem[{Misra} et~al.(2020){Misra}, {Fragos}, {Tauris}, {Zapartas} \&
  {Aguilera-Dena}]{Misra20}
{Misra} D., {Fragos} T., {Tauris} T., {Zapartas} E., {Aguilera-Dena} D.~R.,
  2020, arXiv e-prints,  arXiv:2004.01205

\bibitem[{Mitsuda} et~al.(1984){Mitsuda}, {Inoue}, {Koyama} et~al.]{diskbb}
{Mitsuda} K., {Inoue} H., {Koyama} K., et~al., 1984, \pasj, 36, 741

\bibitem[{Mortlock} et~al.(2011){Mortlock}, {Warren}, {Venemans}
  et~al.]{Mortlock11}
{Mortlock} D.~J., {Warren} S.~J., {Venemans} B.~P., et~al., 2011, \nat, 474,
  616

\bibitem[{Motch} et~al.(2014){Motch}, {Pakull}, {Soria}, {Gris{\'e}} \&
  {Pietrzy{\'n}ski}]{Motch14nat}
{Motch} C., {Pakull} M.~W., {Soria} R., {Gris{\'e}} F., {Pietrzy{\'n}ski} G.,
  2014, \nat, 514, 198

\bibitem[{Mukherjee} et~al.(2015){Mukherjee}, {Walton}, {Bachetti}
  et~al.]{Mukherjee15}
{Mukherjee} E.~S., {Walton} D.~J., {Bachetti} M., et~al., 2015, \apj, 808, 64

\bibitem[{Orosz}(2003)]{Orosz03}
{Orosz} J.~A., 2003, in { A Massive Star Odyssey: From Main Sequence to
  Supernova\/}, edited by {K.~van der Hucht, A.~Herrero, \& C.~Esteban}, vol.
  212 of { IAU Symposium, Astron. Soc. Pac., San Francisco\/},  365

\bibitem[{Parker} et~al.(2019){Parker}, {Buisson}, {Tomsick} et~al.]{Parker19}
{Parker} M.~L., {Buisson} D.~J.~K., {Tomsick} J.~A., et~al., 2019, \mnras, 484,
  1, 1202

\bibitem[{Pinto} et~al.(2017){Pinto}, {Alston}, {Soria} et~al.]{Pinto17}
{Pinto} C., {Alston} W., {Soria} R., et~al., 2017, \mnras, 468, 2865

\bibitem[{Pinto} et~al.(2016){Pinto}, {Middleton} \& {Fabian}]{Pinto16nat}
{Pinto} C., {Middleton} M.~J., {Fabian} A.~C., 2016, \nat, 533, 64

\bibitem[{Pinto} et~al.(2020){Pinto}, {Walton}, {Kara} et~al.]{Pinto20}
{Pinto} C., {Walton} D.~J., {Kara} E., et~al., 2020, \mnras, 492, 4, 4646

\bibitem[{Pintore} et~al.(2018){Pintore}, {Belfiore}, {Novara}
  et~al.]{Pintore18}
{Pintore} F., {Belfiore} A., {Novara} G., et~al., 2018, \mnras, 477, 1, L90

\bibitem[{Pintore} et~al.(2020){Pintore}, {Marelli}, {Salvaterra}
  et~al.]{Pintore20}
{Pintore} F., {Marelli} M., {Salvaterra} R., et~al., 2020, \apj, 890, 2, 166

\bibitem[{Rana} et~al.(2015){Rana}, {Harrison}, {Bachetti} et~al.]{Rana15}
{Rana} V., {Harrison} F.~A., {Bachetti} M., et~al., 2015, \apj, 799, 121

\bibitem[{Ransom} et~al.(2002){Ransom}, {Eikenberry} \&
  {Middleditch}]{PRESTOaccel}
{Ransom} S.~M., {Eikenberry} S.~S., {Middleditch} J., 2002, \aj, 124, 1788

\bibitem[{Reig}(2011)]{Reig11}
{Reig} P., 2011, \apss, 332, 1

\bibitem[{Reynolds} et~al.(2014){Reynolds}, {Reis}, {Miller}, {Cackett} \&
  {Degenaar}]{Reynolds14quies}
{Reynolds} M.~T., {Reis} R.~C., {Miller} J.~M., {Cackett} E.~M., {Degenaar} N.,
  2014, \mnras, 441, 4, 3656

\bibitem[{Riley} et~al.(2019){Riley}, {Watts}, {Bogdanov} et~al.]{Riley19}
{Riley} T.~E., {Watts} A.~L., {Bogdanov} S., et~al., 2019, \apjl, 887, 1, L21

\bibitem[{Roberts} et~al.(2011){Roberts}, {Gladstone}, {Goulding}
  et~al.]{Roberts11}
{Roberts} T.~P., {Gladstone} J.~C., {Goulding} A.~D., et~al., 2011,
  Astronomische Nachrichten, 332, 398

\bibitem[{Rodr{\'\i}guez Castillo} et~al.(2020){Rodr{\'\i}guez Castillo},
  {Israel}, {Belfiore} et~al.]{Rodriguez20}
{Rodr{\'\i}guez Castillo} G.~A., {Israel} G.~L., {Belfiore} A., et~al., 2020,
  \apj, 895, 1, 60

\bibitem[{Sathyaprakash} et~al.(2019){Sathyaprakash}, {Roberts}, {Walton}
  et~al.]{Sathyaprakash19}
{Sathyaprakash} R., {Roberts} T.~P., {Walton} D.~J., et~al., 2019, \mnras, 488,
  1, L35

\bibitem[{Schlafly} \& {Finkbeiner}(2011)]{Schlafly11}
{Schlafly} E.~F., {Finkbeiner} D.~P., 2011, \apj, 737, 2, 103

\bibitem[{Shakura} \& {Sunyaev}(1973)]{Shakura73}
{Shakura} N.~I., {Sunyaev} R.~A., 1973, \aap, 24, 337

\bibitem[{Shimura} \& {Takahara}(1995)]{Shimura95}
{Shimura} T., {Takahara} F., 1995, \apj, 445, 780

\bibitem[{Skinner} et~al.(1982){Skinner}, {Bedford}, {Elsner}, {Leahy},
  {Weisskopf} \& {Grindlay}]{Skinner82}
{Skinner} G.~K., {Bedford} D.~K., {Elsner} R.~F., {Leahy} D., {Weisskopf}
  M.~C., {Grindlay} J., 1982, \nat, 297, 5867, 568

\bibitem[{Song} et~al.(2020){Song}, {Walton}, {Lansbury} et~al.]{Song20}
{Song} X., {Walton} D.~J., {Lansbury} G.~B., et~al., 2020, \mnras, 491, 1, 1260

\bibitem[{Soria} et~al.(2012){Soria}, {Kuntz}, {Winkler} et~al.]{Soria12}
{Soria} R., {Kuntz} K.~D., {Winkler} P.~F., et~al., 2012, \apj, 750, 2, 152

\bibitem[{Steiner} et~al.(2009){Steiner}, {Narayan}, {McClintock} \&
  {Ebisawa}]{simpl}
{Steiner} J.~F., {Narayan} R., {McClintock} J.~E., {Ebisawa} K., 2009, \pasp,
  121, 1279

\bibitem[{Stobbart} et~al.(2006){Stobbart}, {Roberts} \& {Wilms}]{Stobbart06}
{Stobbart} A.-M., {Roberts} T.~P., {Wilms} J., 2006, \mnras, 368, 397

\bibitem[{Str{\"u}der} et~al.(2001){Str{\"u}der}, {Briel}, {Dennerl}
  et~al.]{XMM_PN}
{Str{\"u}der} L., {Briel} U., {Dennerl} K., et~al., 2001, \aap, 365, L18

\bibitem[{Swartz} et~al.(2011){Swartz}, {Soria}, {Tennant} \&
  {Yukita}]{Swartz11}
{Swartz} D.~A., {Soria} R., {Tennant} A.~F., {Yukita} M., 2011, \apj, 741, 49

\bibitem[{Tao} et~al.(2019){Tao}, {Feng}, {Zhang} et~al.]{Tao19}
{Tao} L., {Feng} H., {Zhang} S., et~al., 2019, \apj, 873, 1, 19

\bibitem[{Tao} et~al.(2012){Tao}, {Kaaret}, {Feng} \& {Gris{\'e}}]{Tao12}
{Tao} L., {Kaaret} P., {Feng} H., {Gris{\'e}} F., 2012, \apj, 750, 110

\bibitem[{Tsygankov} et~al.(2016){Tsygankov}, {Mushtukov}, {Suleimanov} \&
  {Poutanen}]{Tsygankov16}
{Tsygankov} S.~S., {Mushtukov} A.~A., {Suleimanov} V.~F., {Poutanen} J., 2016,
  \mnras, 457, 1101

\bibitem[{Turner} et~al.(2001){Turner}, {Abbey}, {Arnaud} et~al.]{XMM_MOS}
{Turner} M.~J.~L., {Abbey} A., {Arnaud} M., et~al., 2001, \aap, 365, L27

\bibitem[{van Haaften} et~al.(2019){van Haaften}, {Maccarone}, {Rhode}, {Kundu}
  \& {Zepf}]{vanHaaften19}
{van Haaften} L.~M., {Maccarone} T.~J., {Rhode} K.~L., {Kundu} A., {Zepf}
  S.~E., 2019, \mnras, 483, 3, 3566

\bibitem[{van Paradijs} \& {McClintock}(1994)]{Vanparadijs94}
{van Paradijs} J., {McClintock} J.~E., 1994, \aap, 290, 133

\bibitem[{Vasilopoulos} et~al.(2020{\natexlab{a}}){Vasilopoulos}, {Lander},
  {Koliopanos} \& {Bailyn}]{Vasilopoulos20}
{Vasilopoulos} G., {Lander} S.~K., {Koliopanos} F., {Bailyn} C.~D.,
  2020{\natexlab{a}}, \mnras, 491, 4, 4949

\bibitem[{Vasilopoulos} et~al.(2019){Vasilopoulos}, {Petropoulou}, {Koliopanos}
  et~al.]{Vasilopoulos19}
{Vasilopoulos} G., {Petropoulou} M., {Koliopanos} F., et~al., 2019, \mnras,
  488, 4, 5225

\bibitem[{Vasilopoulos} et~al.(2020{\natexlab{b}}){Vasilopoulos}, {Ray},
  {Gendreau} et~al.]{Vasilopoulos20rx}
{Vasilopoulos} G., {Ray} P.~S., {Gendreau} K.~C., et~al., 2020{\natexlab{b}},
  \mnras, 494, 4, 5350

\bibitem[{Verbunt}(1993)]{Verbunt93rev}
{Verbunt} F., 1993, \araa, 31, 93

\bibitem[{Verner} et~al.(1996){Verner}, {Ferland}, {Korista} \&
  {Yakovlev}]{Verner96}
{Verner} D.~A., {Ferland} G.~J., {Korista} K.~T., {Yakovlev} D.~G., 1996, \apj,
  465, 487

\bibitem[{Walton} et~al.(2018{\natexlab{a}}){Walton}, {Bachetti}, {F{\"u}rst}
  et~al.]{Walton18crsf}
{Walton} D.~J., {Bachetti} M., {F{\"u}rst} F., et~al., 2018{\natexlab{a}},
  \apjl, 857, L3

\bibitem[{Walton} et~al.(2016{\natexlab{a}}){Walton}, {F{\"u}rst}, {Bachetti}
  et~al.]{Walton16period}
{Walton} D.~J., {F{\"u}rst} F., {Bachetti} M., et~al., 2016{\natexlab{a}},
  \apjl, 827, L13

\bibitem[{Walton} et~al.(2017){Walton}, {F{\"u}rst}, {Harrison}
  et~al.]{Walton17hoIX}
{Walton} D.~J., {F{\"u}rst} F., {Harrison} F.~A., et~al., 2017, \apj, 839, 105

\bibitem[{Walton} et~al.(2018{\natexlab{b}}){Walton}, {F{\"u}rst}, {Harrison}
  et~al.]{Walton18p13}
{Walton} D.~J., {F{\"u}rst} F., {Harrison} F.~A., et~al., 2018{\natexlab{b}},
  \mnras, 473, 4360

\bibitem[{Walton} et~al.(2018{\natexlab{c}}){Walton}, {F{\"u}rst}, {Heida}
  et~al.]{Walton18ulxBB}
{Walton} D.~J., {F{\"u}rst} F., {Heida} M., et~al., 2018{\natexlab{c}}, \apj,
  856, 128

\bibitem[{Walton} et~al.(2011{\natexlab{a}}){Walton}, {Gladstone}, {Roberts}
  et~al.]{Walton4517}
{Walton} D.~J., {Gladstone} J.~C., {Roberts} T.~P., et~al., 2011{\natexlab{a}},
  \mnras, 414, 1011

\bibitem[{Walton} et~al.(2015{\natexlab{a}}){Walton}, {Harrison}, {Bachetti}
  et~al.]{Walton15}
{Walton} D.~J., {Harrison} F.~A., {Bachetti} M., et~al., 2015{\natexlab{a}},
  \apj, 799, 122

\bibitem[{Walton} et~al.(2014){Walton}, {Harrison}, {Grefenstette}
  et~al.]{Walton14hoIX}
{Walton} D.~J., {Harrison} F.~A., {Grefenstette} B.~W., et~al., 2014, \apj,
  793, 21

\bibitem[{Walton} et~al.(2016{\natexlab{b}}){Walton}, {Middleton}, {Pinto}
  et~al.]{Walton16ufo}
{Walton} D.~J., {Middleton} M.~J., {Pinto} C., et~al., 2016{\natexlab{b}},
  \apjl, 826, L26

\bibitem[{Walton} et~al.(2015{\natexlab{b}}){Walton}, {Middleton}, {Rana}
  et~al.]{Walton15hoII}
{Walton} D.~J., {Middleton} M.~J., {Rana} V., et~al., 2015{\natexlab{b}}, \apj,
  806, 65

\bibitem[{Walton} et~al.(2020){Walton}, {Pinto}, {Nowak} et~al.]{Walton20}
{Walton} D.~J., {Pinto} C., {Nowak} M., et~al., 2020, \mnras, 494, 4, 6012

\bibitem[{Walton} et~al.(2011{\natexlab{b}}){Walton}, {Roberts}, {Mateos} \&
  {Heard}]{WaltonULXcat}
{Walton} D.~J., {Roberts} T.~P., {Mateos} S., {Heard} V., 2011{\natexlab{b}},
  \mnras, 416, 1844

\bibitem[{Walton} et~al.(2016{\natexlab{c}}){Walton}, {Tomsick}, {Madsen}
  et~al.]{Walton16cyg}
{Walton} D.~J., {Tomsick} J.~A., {Madsen} K.~K., et~al., 2016{\natexlab{c}},
  \apj, 826, 87

\bibitem[{Watarai} \& {Mineshige}(2003)]{Watari03}
{Watarai} K.-y., {Mineshige} S., 2003, \apj, 596, 1, 421

\bibitem[{Webb} et~al.(2020){Webb}, {Coriat}, {Traulsen} et~al.]{4XMM}
{Webb} N.~A., {Coriat} M., {Traulsen} I., et~al., 2020, arXiv e-prints,
  arXiv:2007.02899

\bibitem[{Weisskopf} et~al.(2002){Weisskopf}, {Brinkman}, {Canizares},
  {Garmire}, {Murray} \& {Van Speybroeck}]{CHANDRA}
{Weisskopf} M.~C., {Brinkman} B., {Canizares} C., {Garmire} G., {Murray} S.,
  {Van Speybroeck} L.~P., 2002, \pasp, 114, 1

\bibitem[{Wilms} et~al.(2000){Wilms}, {Allen} \& {McCray}]{tbabs}
{Wilms} J., {Allen} A., {McCray} R., 2000, \apj, 542, 914

\bibitem[{Wilson-Hodge} et~al.(2018){Wilson-Hodge}, {Malacaria}, {Jenke}
  et~al.]{WilsonHodge18}
{Wilson-Hodge} C.~A., {Malacaria} C., {Jenke} P.~A., et~al., 2018, \apj, 863,
  1, 9

\bibitem[{Zampieri} \& {Roberts}(2009)]{Zampieri09}
{Zampieri} L., {Roberts} T.~P., 2009, \mnras, 400, 677

\end{thebibliography}

\label{lastpage}

\end{document}